\documentclass[sigconf]{acmart}

\usepackage{enumitem}
\usepackage{booktabs}
\usepackage{multirow}
\usepackage{makecell}
\usepackage{tabularx}
\usepackage{graphicx} 
\usepackage{longtable}
\usepackage{float}
\usepackage{arydshln}
\usepackage[most]{tcolorbox}

\AtBeginDocument{%
  }

\setcopyright{acmlicensed}
\copyrightyear{2026}
\acmYear{2026}
\acmDOI{XXXXXXX.XXXXXXX}
\acmConference[Conference acronym 'XX]{Make sure to enter the correct
  conference title from your rights confirmation email}{June 03--05,
  2018}{Woodstock, NY}
\acmISBN{978-1-4503-XXXX-X/2018/06}

\newcommand{\system}{MAP}

\newcommand{\ui}[1]{\ensuremath{\langle \textnormal{#1} \rangle}}

\begin{document}

\title{Comparing Computer Use Behavior Across Multiple Levels of Abstraction}


\author{\mbox{Ananya Gubbi Mohanbabu}}
\affiliation{%
  \institution{\mbox{University of California, Berkeley}}
  \city{Berkeley, CA}
  \country{USA}}
\email{ananyagm@berkeley.edu}

\author{Rosiana Natalie}
\affiliation{%
  \institution{University of Michigan}
  \city{Ann Arbor, MI}
  \country{USA}}
\email{rosianan@umich.edu}

\author{Anhong Guo}
\affiliation{%
  \institution{University of Michigan}
  \city{Ann Arbor, MI}
  \country{USA}}
\email{anhong@umich.edu}

\author{Amy Pavel}
\affiliation{%
  \institution{\mbox{University of California, Berkeley}}
  \city{Berkeley, CA}
  \country{USA}}
\email{amypavel@berkeley.edu}




\begin{teaserfigure}      
\includegraphics[width=\textwidth]{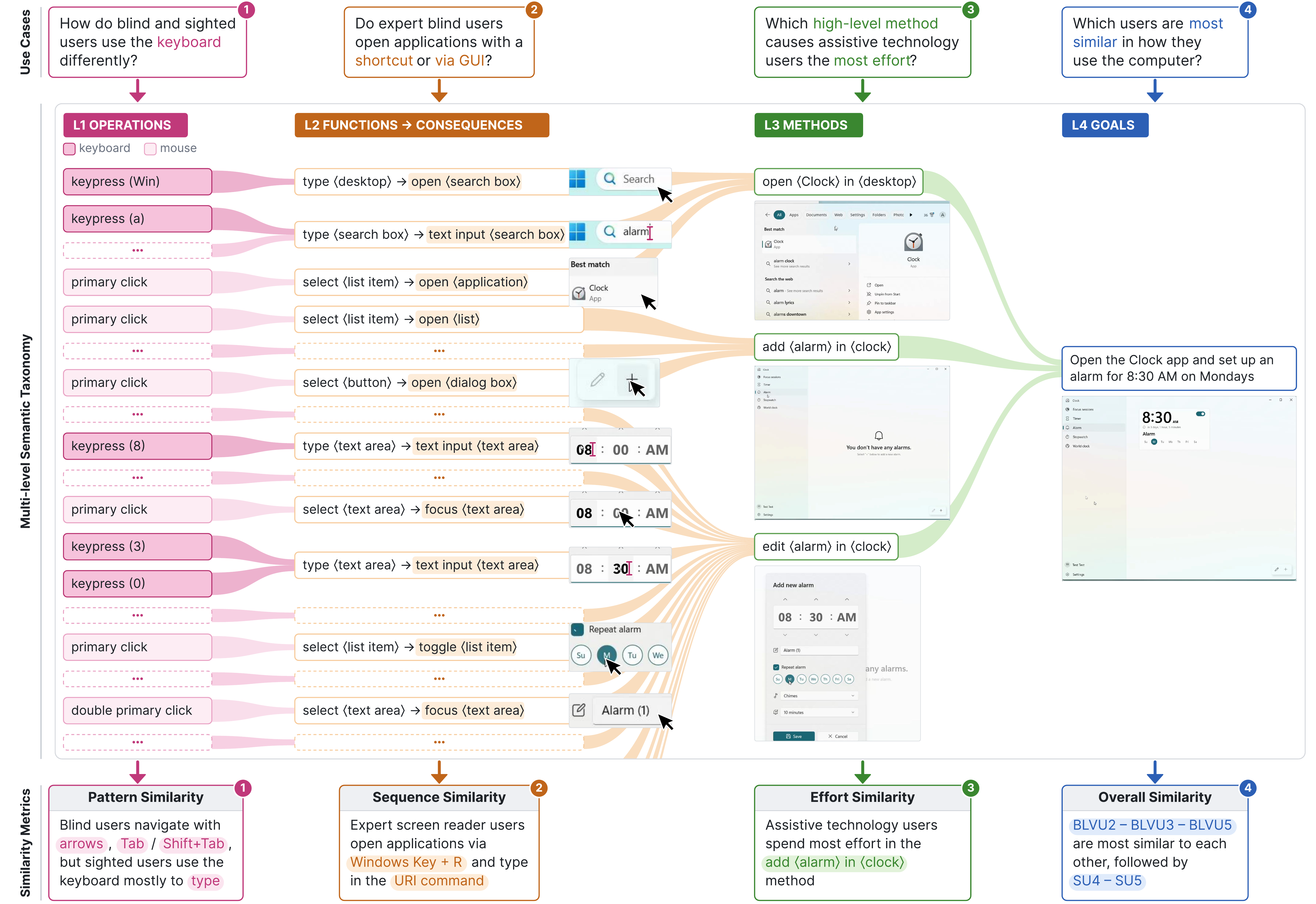}
  \centering
  \Description{Overview of MAP’s multi-level semantic representation and comparison metrics using an alarm-setting task. The center shows one interaction trace transformed from L1 Operations, such as keypresses, clicks, and pointing, into L2 Functions and Consequences, then L3 Methods such as opening the Clock application and adding an alarm, and finally the L4 Goal, “Set up an alarm for 8:30 AM on Mondays.” Curved links show how lower-level actions contribute to higher-level labels. Callouts above illustrate questions that can be asked at different abstraction levels, such as which interaction behaviors are common among blind users or which Methods require the most effort. Callouts below illustrate MAP’s three similarity metrics: Pattern Similarity identifies recurring behaviors, Sequence Similarity compares the order of interactions, and Effort Similarity compares where users spend actions and time.}
  \caption{\system{} transforms raw interaction traces into multi-level semantic representations (center), enabling comparison of user behavior from low-level actions to high-level tasks. Given two traces, \system{} supports comparisons of similarity between users (top) based on sequence alignment, effort allocation, and recurring behavioral patterns (bottom).}
  \label{fig:teaser}
\end{teaserfigure}


\begin{abstract}
Interaction traces record how people and agents use computers, and comparing them can reveal usability barriers, differences in expertise, and human--agent similarity. Yet traces are recorded as low-level, interface-specific actions, making comparisons difficult across tasks, interfaces, and input modalities. Existing approaches either compare these actions directly, reduce traces to aggregate outcomes, or summarize them at higher-level goals, obscuring where behavior differs. We present \system{} (Multi-level Action Parsing), a  hierarchical semantic representation and annotation pipeline that maps traces into a common vocabulary across linked levels of abstraction. \system{} compares traces by sequence, recurring patterns, and effort, localizing differences by type and abstraction level. A technical evaluation across three computer use datasets shows high agreement with manual annotations and broad coverage of behavior. Applying \system{} shows that sighted and blind users increasingly converge at higher levels despite navigation differences, while agents omit corrective, help-seeking, and recovery behaviors present in human traces.
\end{abstract}

\maketitle

\section{Introduction}
Understanding how people use computers has long been a core goal of HCI~\cite{hilbert2000extracting}. A rich source of evidence is the \textit{interaction trace}: a logged sequence of computer use actions (\textit{e.g.}, clicks and keypresses) together with their context (\textit{e.g.}, applications and interface elements)~\cite{hilbert2000extracting, zhang2023exploring}. Researchers use such traces to understand how an individual performed a task and to \textit{compare} how people use computers --- \textit{e.g.}, to reveal usability issues by identifying deviations from an ideal trace~\cite{waterson2002did, siochi1991computer}, characterize interaction patterns of experts and novices~\cite{grossman2015investigation}, and surface web browsing strategies~\cite{wang2016unsupervised, clark2006combining}. 

Prior work makes such comparisons by holding the task, interface, and input modality constant to compare low-level actions directly~\cite{waterson2002did, siochi1991computer, wang2016unsupervised}, or by reducing the trace to aggregate measures such as completion time, action count, or success~\cite{grossman2015investigation, abhyankar2026osworld, mohanbabu2026a11y}. But, such work does not yet compare computer use behaviors that cut \textit{across} tasks, interfaces, and input modalities -- \textit{e.g.} to compare across tasks to surface usability barriers, across modalities to locate accessibility gaps, or across interfaces to characterize general computer use expertise. Recent datasets of interaction traces record this kind of variation~\cite{deng2023mind2web, rawles2023androidinthewild, xie2024osworld, mohanbabu2026a11y}, creating the opportunity to make such comparisons at scale.

But, it remains challenging to compare traces as they are recorded in a low-level input- and interface-specific vocabulary. 
For example, a screen reader user may press Tab six times while a sighted user points to reach the same button, but their traces differ.
Recent work thus describes traces at a higher level of abstraction by using LLMs to summarize low-level actions into goals (\textit{e.g.}, search for a product, fill in the shipping form)~\cite{zhang2025summact, wang2025ai}, then aligns those descriptions to report where two traces cover the same steps~\cite{wang2025ai}.
However, as each trace is labeled \textit{independently} the same goal may be described in different ways, such that it remains challenging to count or match goals across a corpus of traces (\textit{e.g.}, to identify where a group of users got stuck). 
Further, this level of abstraction is useful for comparing higher-level workflows but obscures differences we care about when comparing how people use computers: people can reach the same goal by different methods (\textit{e.g.}, search vs. faceted browsing), or follow the same method with different costs (\textit{e.g.}, navigating facets linearly vs. by heading).
To meaningfully compare computer use behavior across heterogeneous interaction traces thus requires a representation with a \textit{shared vocabulary} to aggregate labels across traces and \textit{multiple linked levels of abstraction} so that we can recover \textit{how} and \textit{at what level} interaction traces differ.

We introduce \textbf{\system{}} (Multi-level Action Parsing), a framework for comparing computer use behavior across multiple levels of abstraction and dimensions of similarity. \system{} provides a hierarchical semantic representation of interaction traces across four connected levels, from input actions as \textbf{Operations (L1)}, to their \textbf{Functions \& Consequences (L2)}, \textbf{Methods (L3)}, and \textbf{Goals (L4)}. 
We construct these levels through the following procedure: \textbf{Operations (L1)} are low-level actions common to interaction traces (\textit{e.g.}, pointing, clicking), \textbf{Functions and Consequences (L2)} are derived by unifying existing standards that describe how interfaces are implemented~\cite{development-guide-for-the-accessibility-infrastructure-documentation_2023, karl-bridge-microsoft_2025} (\textit{e.g.}, ``focus $\langle$search box$\rangle$''), \textbf{Methods (L3)} are derived by open-coding interaction traces and consolidating equivalent labels (\textit{e.g.}, ``edit'', ``search''), and \textbf{Goals (L4)} are task-dependent and thus open-vocabulary as in prior work (\textit{e.g.}, ``Edit the file named final.txt'')~\cite{wang2025ai, zhang2025summact}. \system{}'s vocabulary instantiates this procedure for desktop GUI computer use. 
To assign this vocabulary to interaction traces at scale, \system{} applies labels deterministically when possible (\textit{e.g.}, L1 operations and L2 functions) and otherwise with an LLM with a fixed vocabulary (\textit{e.g.}, L3 methods), such that only task-dependent fields remain open-vocabulary.
\system{}’s shared representation thus makes it possible to identify where traces diverge.

To capture different ways in which behavior can be similar, \system{} applies three complementary metrics: \textbf{sequence similarity}, which measures alignment in the order of interactions; \textbf{pattern similarity}, which captures recurring local interaction patterns; and \textbf{effort similarity}, which captures how actions and time are allocated across a workflow. Each metric can be computed at each abstraction level, enabling \system{} to localize both the \textit{type} and \textit{level} of behavioral differences between interaction traces.

We validate \system{} via a technical evaluation and an examination of its capacity to answer questions posed in prior research. 
Our technical evaluation compares \system{}'s annotations against manual annotations of 2,342 actions across three computer use datasets spanning human and agent traces and sighted and screen reader interactions (A11y-CUA~\cite{mohanbabu2026a11y}, AgentSynth~\cite{xie2026agentsynth}, and AgentNet~\cite{yang2026agentnet}). 
\system{} agrees with manual annotation on 77.3-87.3\% of actions per semantic field, and 99.2\% of actions across the full corpus receive a defined label rather than \textit{unknown}, indicating that the vocabularies cover observed behavior. We further characterize where the pipeline disagrees with annotators: most L3 errors reflect segmentation differences, and L2 errors rarely propagate to L3. 

We use \system{} to replicate and extend prior findings on how sighted users, blind users, and computer use agents complete the same tasks~\cite{sahib2012comparative,borodin2010more,mohanbabu2026a11y,wang2025ai}. Consistent with prior work, blind and sighted users differ sharply in low-level interactions, and agents differ from their corresponding human groups in how they execute tasks. \system{} further localizes where these differences occur: the additional effort for blind users is concentrated in navigation rather than selection or typing, while the two groups largely converge on the same higher-level Methods. Agents likewise become more similar to humans at the Method level, but reproduce fewer of the corrective, help-seeking, and recovery behaviors found in human traces.
We then discuss opportunities to apply \system{}'s framework in the future to \textbf{surface efficient strategies} from expert traces to inform novice users on how to improve, to \textbf{localize usability problems} on interfaces across tasks and users, and to \textbf{share a user's computer use behavior} without exposing their identity. We also discuss limitations of \system{} and potential opportunities to generalize \system{} beyond desktop GUIs in the future. 
Collectively, \system{} enables researchers to compare computer use behavior across tasks, interfaces, and input modalities by localizing similarities and differences at multiple abstraction levels.

In summary, we contribute:

\begin{itemize}

\item \textbf{\system{}, a hierarchical semantic representation of interaction traces} that labels recorded actions across four levels of abstraction and an automatic pipeline that assigns these labels to interaction traces.

\item \textbf{A decomposition of behavioral similarity into three dimensions} --- sequence, recurring patterns, and effort allocation --- computed with standard measures over \system{} labels at any level, so that behavioral differences can be localized by type and level of abstraction.

\item \textbf{A technical evaluation and comparative analysis using \system{}} --- replicating and extending prior work comparing blind and sighted computer users, and assessing differences between human and agent traces.
\end{itemize}

\section{Related Work}

Interaction traces capture how humans and agents carry out computer tasks, from individual inputs to higher-level goals. Prior work has recorded and represented traces in machine-readable formats, abstracted interaction traces into meaningful higher-level units, and compared executions with interaction traces. We build on these directions to create \system{}, a shared semantic hierarchy that captures intermediate functional levels between inputs and goals, enabling consistent analysis across traces.

\subsection{Recording Interaction Traces}

An interaction trace records a sequence of low-level actions a person or agent performs while completing a computer task, often together with the interface state in which each action occurred. Such traces are used to study computer use behavior~\cite{mohanbabu2026a11y} and to train and evaluate computer use agents~\cite{rawles2023androidinthewild,deng2023mind2web,xie2024osworld,wang2026opencua}. Prior work has progressively enriched these records with information about the context of each action~\cite{deka2016erica,deka2017rico,abb2024process}. Early systems paired actions with UI states~\cite{deka2016erica}, while later datasets added screenshots and structured interface hierarchies~\cite{deka2017rico}. More recent work standardizes which fields a UI log should record, including actions, UI elements and states, applications, and tasks, so logs from different collection tools can be combined for analysis and automation~\cite{abb2024process}. This standardizes what events are recorded, but not how different events that accomplish the same behavior are represented. For example, clicking a \texttt{Save} button and pressing \texttt{Ctrl+S} can be recorded using the same fields, but the data model does not itself represent that both actions accomplish \texttt{save}. \system{} addresses this second problem by mapping different recorded actions to shared semantic labels based on what they accomplish.

OneTrace similarly brings interaction histories from different applications and users into a common trace structure so that they can be viewed and used together~\cite{battut2025onetrace}. Users then manually group these low-level interaction traces they consider relevant to the same activity. \system{} differs by automatically mapping heterogeneous actions to shared semantic labels based on what they accomplish and the Methods they contribute to. Thus, mouse pointing and repeated keyboard navigation can remain distinct low-level actions while being recognized as equivalent at a higher semantic level.

Interaction traces support learning, behavioral analysis, and agent evaluation and debugging~\cite{li2021screen2vec,he2021actionbert,rawles2023androidinthewild,deng2023mind2web,xie2024osworld,wang2026opencua,mohanbabu2026a11y,desmond2025agent,wang2026survey}. Learned representations of screens and UI components support retrieval and action prediction~\cite{li2021screen2vec,he2021actionbert}; computer use datasets use traces to train agents, evaluate task completion, and study behavior~\cite{rawles2023androidinthewild,deng2023mind2web,xie2024osworld,wang2026opencua,mohanbabu2026a11y}; and tools use traces to inspect agent behavior or identify failures~\cite{desmond2025agent,wang2026survey}. These analyses generally operate on recorded actions, interface states, or task outcomes, making direct comparison difficult when different inputs accomplish equivalent behavior. \system{} preserves those low-level differences while additionally mapping them into shared Functions, Consequences, Methods, and Goals, allowing traces to be compared at the level where their behavior becomes equivalent.

\subsection{Semantic Abstraction of Interaction Traces}

Abstracting interaction traces makes it possible to recognize when different low-level actions accomplish the same thing and to compare behavior beyond exact input sequences. Task models provide an early precedent for representing computer use at multiple levels of abstraction. Hierarchical Task Analysis decomposes tasks into goals and subgoals~\cite{stanton2006hierarchical}, GOMS represents goals through methods and lower-level operators~\cite{john1996goms}, and the Keystroke-Level Model represents execution through primitive operators~\cite{card1980keystroke}. These models are typically constructed top-down by researchers for a known task rather than inferred from recorded traces. \system{} adopts this hierarchical structure but derives its labels from traces, connecting low-level Operations (L1) to what actions do and change (L2), Methods (L3), and task-level Goals (L4).

More recent work infers semantic structure directly from recorded traces. Early systems classified sequences of UI states into predefined procedure-like categories, such as searching or composing~\cite{deka2016erica}, corresponding to \system{}'s Methods (L3). Like \system{}, these categories are drawn from a fixed vocabulary. However, they represent only this procedure level rather than linking it to lower-level actions, their outcomes, and higher-level task Goals. LLM-based approaches instead infer open-ended descriptions: some extract executable multi-action units such as logging in or booking a flight~\cite{huang2024automatic}, while others group low-level actions into intermediate subgoals and then an overall task goal~\cite{zhang2025summact}. Recent work similarly induces natural-language steps from human and agent traces and aligns them to compare how the same task is completed~\cite{wang2025ai}. These approaches capture semantics similar to \system{}'s Methods and Goals, but their labels are generated independently for each trace, so equivalent behavior may be phrased differently and must be matched afterward. They also do not represent \system{}'s L2 distinction between what an action does, what it acts on, and what changes as a result. \system{} instead uses shared vocabularies at L2 and L3 and links them to both the underlying Operations and task-dependent Goals, enabling equivalent behavior to be aggregated across traces while preserving how it was carried out.

\subsection{Comparing User Workflows}

Prior work compares low-level interaction traces directly. Clickstream visualizations reveal common page transitions~\cite{zhao2015matrixwave}, while sequential pattern mining identifies ordered action subsequences that recur across traces~\cite{liu2016patterns}. Other tools align traces around selected actions~\cite{krause2023visually} or compare observed traces with supported task paths to identify inefficient behavior~\cite{filosa2025travis}. These approaches work well when traces share the same interface and action vocabulary, but equivalent behavior may appear different when users accomplish the same thing through different low-level actions.

Other work first transforms low-level traces into representations specific to a task or agent environment~\cite{chung2026did,wu2026agenttrails}. Traces are compared using higher-level semantics by defining a semantic model for each benchmark website that maps GUI interactions to task-level actions and states~\cite{chung2026did}. This enables precise comparison within that environment, but the semantic actions must be defined separately for each website and do not directly provide a shared representation across different applications or interfaces. Tool-using agent traces can instead be aligned through tool calls and dependencies among their inputs, outputs, and artifacts~\cite{wu2026agenttrails}. This requires a shared tool environment rather than providing a common semantic representation of computer use behavior. \system{} instead applies the same semantic representation across tasks, interfaces, and input modalities, allowing heterogeneous traces to be compared without defining environment-specific semantic actions or requiring shared tools.
\section{\system{} Taxonomy}
\label{sec:taxonomy}

To enable systematic comparison of computer use interaction traces across multiple users and tasks, we introduce \system{}, a hierarchical labeling approach to semantically annotate traces from various domains including web browsing, system operations, media, document editing, and workflow applications. \system{} represents interaction traces at four connected levels so behavior can be compared at different granularities. Operations (L1) capture the recorded input, Functions \& Consequences (L2) capture what each action does and changes in the interface, Methods (L3) capture how users accomplish an immediate objective, and Goals (L4) capture the broader task outcome. 
The hierarchy builds on the decomposition used in GOMS~\cite{john1996goms}, while adding interface-level semantics between L1 operators and L3 methods (Table~\ref{tab:map-abstraction-levels} summarizes each level, refer to Appendix ~\ref{sec:vocabulary} for full taxonomy).

To support comparisons across users, tasks, and datasets, we aimed for the \system{} taxonomy to provide \textbf{coverage} for interactions across diverse interfaces, \textbf{generalize} across applications and operating systems, and assign \textbf{consistent} labels to behaviors so recurring patterns can be aggregated within and across traces. \system{} therefore annotates behavior from the recorded actions and their interface context, including the active application, interacted UI elements, and assistive technology when needed.

\begin{table*}[t]
\centering
\small
\setlength{\tabcolsep}{2.5pt}
\renewcommand{\arraystretch}{1.2}

\begin{tabular}{
    p{0.025\textwidth}
    p{0.11\textwidth}
    p{0.31\textwidth}
    p{0.30\textwidth}
    p{0.19\textwidth}
}
\toprule
\textbf{L\#} &
\textbf{Level} &
\textbf{Definition} &
\textbf{Examples} &
\textbf{Related Work} \\
\midrule

L1 &
Operations &
Atomic input actions executed through mouse or keyboard. &
\texttt{primary click}, \texttt{scroll}, \texttt{ keypress}, \texttt{shortcut} &
Keystroke-Level Model~\cite{card1980keystroke} \\

L2 &
Functions \& Consequences &
Captures what an operation does, which UI element it acts on, and the resulting change to the interface. &
\begin{tabular}[t]{@{}l@{}}
\texttt{select} $\langle$button$\rangle$
$\rightarrow$
\texttt{open} $\langle$dialog box$\rangle$ \\[2pt]
\texttt{navigate} $\langle$link$\rangle$
$\rightarrow$
\texttt{focus} $\langle$link$\rangle$
\end{tabular}
&

AT-SPI2~\cite{development-guide-for-the-accessibility-infrastructure-documentation_2023},
UIA Control Patterns~\cite{karl-bridge-microsoft_2025}, WAI-ARIA,
Keyboard shortcuts~\cite{keyboard-shortcuts-in-apps-microsoft-support_2025,Apple-Support} \\

L3 &
Methods \& UI Containers &
Captures how a sub-goal is carried out within an application. The same method may be realized through different sequences of operations and functions. &
\begin{tabular}[t]{@{}l@{}}
\texttt{edit} $\langle$paragraph$\rangle$  in $\langle${text document}$\rangle$ \\[2pt]
\texttt{add} $\langle$chart$\rangle$  in  $\langle${spreadsheet}$\rangle$
\end{tabular}
&
GOMS~\cite{john1996goms} \\

L4 &
Goals &
Captures the intended task outcome. A goal may span one or more applications. &
\begin{tabular}[t]{@{}l@{}}
Export spreadsheet as PDF \\[2pt]
Schedule a meeting
\end{tabular}
&
~\cite{wang2025ai, zhang2025summact} \\

\bottomrule
\end{tabular}

\caption{
The four semantic abstraction levels in \system{}, with each level's definition, example labels, and related work that informed its design.
L1 represents individual input operations.
L2 represents interaction semantics as
{Function} $\langle${Direct UI Target}$\rangle
\rightarrow$ {Consequence} $\langle${Indirect UI Target}$\rangle$.
L3 represents methods as {Method} $\langle${Specific Entity}$\rangle$ in $\langle${UI Container}$\rangle$,
and L4 captures the goal those interactions accomplish. }
\Description{The four semantic abstraction levels in \system{}, with each level's definition, example labels, and related work that informed its design.
L1 represents individual input operations.
L2 represents interaction semantics as
{Function} $\langle${Direct UI Target}$\rangle
\rightarrow$ {Consequence} $\langle${Indirect UI Target}$\rangle$.
L3 represents methods as {Method} $\langle${Specific Entity}$\rangle$ in $\langle${UI Container}$\rangle$,
and L4 captures the goal those interactions accomplish. }
\label{tab:map-abstraction-levels}
\end{table*}

\subsection{Operations (L1)} 

Operations preserve the low-level mouse or keyboard input used to interact with the interface, allowing \system{} to compare how users physically carry out the same behavior before higher-level abstraction. \system{} represents mouse input as \texttt{scroll}, \texttt{primary click}, \texttt{double primary click}, \texttt{secondary click}, \texttt{point}, and \texttt{drag drop}, and keyboard input as \texttt{keypress} or \texttt{shortcut} (Table~\ref{tab:operation-function-mapping}). Because datasets record these inputs differently (\textit{e.g.}, \texttt{mouse\_press} vs.\ \texttt{mouse\_click}, or separate \texttt{mouse\_down} and \texttt{mouse\_up} events), they need to be mapped to this common L1 operation set. We also collapse consecutive \texttt{point} and \texttt{scroll} events, since their logging frequency varies across datasets and would otherwise reflect sampling differences rather than behavior. Cursor coordinates are retained, when available, as context for assigning higher-level labels.

Keyboard input requires additional normalization because the same raw key event may represent ordinary keypress or a shortcut, particularly for screen reader interaction. We therefore use a dictionary of approximately 400 known shortcuts, including screen reader commands, and their expected outcomes to deterministically distinguish \texttt{shortcut} from \texttt{keypress} operations (Table~\ref{tab:keyboard-operation-function-mapping}).

\begin{table}[t]
\centering
\small
\begin{tabular}{p{0.13\linewidth} p{0.35\linewidth} p{0.25\linewidth} p{0.1\linewidth}}
\toprule
\textbf{Operation} & \textbf{Intent} & \textbf{Examples} & \textbf{Function} \\
\midrule

\texttt{Keypress}
& Printable text entered in an editable UI element
& \texttt{a}, \texttt{5}, \texttt{h} in a text field
& Type \\
& UI navigation
& \texttt{Tab}, $\downarrow$, \texttt{PageDown}
& Navigate \\
& Activates or selects the focused UI element
& \texttt{Enter}, \texttt{Space} on a button
& Select \\
\midrule
\texttt{Shortcut}
& Printable text entered in an editable UI element
& \texttt{Shift+h} (for "\texttt{H}")
& Type \\
& UI navigation, including screen-reader commands
& \texttt{Shift+h} (with screen reader ON), \texttt{Ctrl+Tab}, \texttt{Alt+Tab}
& Navigate \\
& Activates or selects the focused UI element
& \texttt{NVDA+Enter}, \texttt{VO+Space}
& Select \\
& Invokes an application- or OS-level command
& \texttt{Ctrl+S}, \texttt{Ctrl+C}, \texttt{F1}-\texttt{F12}
& Program \\

\bottomrule
\end{tabular}
\caption{Mapping L1 keyboard operations to L2 functions. L1 keyboard operations are classified as \texttt{keypress} or \texttt{shortcut} based on the executed key(s) and known shortcut mappings, while their L2 function is determined from the interaction context.}
\Description{Table mapping L1 keyboard Operations to L2 Functions based on interaction intent. A keypress can represent typing printable text, navigating the interface with keys such as Tab or PageDown, or selecting the focused element with Enter or Space. A shortcut can similarly type text, navigate - including screen-reader commands such as Shift+H -select the focused element, or invoke an application or operating-system command such as Ctrl+S, Ctrl+C, or function keys. The resulting L2 Functions are Type, Navigate, Select, or Program, determined from the interaction context rather than from the key alone.}
\label{tab:keyboard-operation-function-mapping}
\end{table}

\subsection{Functions \& Consequences (L2)}

L2 makes individual actions comparable by describing what each action does in the interface and what changes as a result. Each action is represented by its \textit{function} and the UI element it acts on, followed by its \textit{consequence} and the UI element where that change occurs (if applicable). For example, selecting a button may close a dialog box:
\texttt{select} \ui{button} $\rightarrow$ \texttt{close} \ui{dialog box}.

Functions describe how an L1 operation acts on its direct UI target. Mouse operations map deterministically to functions (Table~\ref{tab:operation-function-mapping}): for example, \texttt{primary click} maps to \texttt{select}, while \texttt{point} maps to \texttt{navigate}. Keyboard operations depend on context because the same key can serve different functions. For example, \texttt{Shift+h} may type ``H'' in a text field or navigate to a heading when a screen reader is active. We therefore use the focused UI element and screen-reader state to deterministically distinguish typing, navigation, selection, and program commands (Table~\ref{tab:keyboard-operation-function-mapping}).

Consequences describe the observable change produced by an action and where that change occurs. In
\texttt{select} \ui{button} $\rightarrow$ \texttt{close} \ui{dialog box},
the button is acted on directly, while the dialog box is where the resulting change occurs. Separating an action from its outcome distinguishes interactions that perform the same function on the same type of element but produce different interface changes.

We derive fixed vocabularies for functions, consequences, and UI targets from existing interface and accessibility frameworks, including WAI-ARIA, AT-SPI, and UI Automation Control Patterns~\cite{development-guide-for-the-accessibility-infrastructure-documentation_2023, keyboard-shortcuts-in-apps-microsoft-support_2025, Apple-Support}. We consolidate semantically equivalent concepts across frameworks. For example, \texttt{SelectionItemPattern.Select()}~\cite{karl-bridge-microsoft_2025}, \texttt{STATE\_SELECTED}~\cite{development-guide-for-the-accessibility-infrastructure-documentation_2023}, and \texttt{aria-selected}~\cite{development-guide-for-the-accessibility-infrastructure-documentation_2023} map to \texttt{select}. An LLM assigns UI targets and consequences from these fixed vocabularies (Appendix ~\ref{sec:consequences}, ~\ref{sec:ui-targets}) using the interaction trace and preceding labels (Table~\ref{tab:map-inference-source}), and it assigns \texttt{Unknown} when the trace provides insufficient evidence for any of the fixed labels in our vocabulary.

L2 captures what occurred in the interface, but not why it contributed to the user's task. For example, \texttt{select} \ui{check box} could correspond to filtering products or completing a form. L3 captures this higher-level purpose.

\begin{table}[]
\small
\centering
\begin{tabular}{llp{0.24\linewidth}}
\toprule
\textbf{L1 Operations} & \textbf{L1 Specific Details} & \textbf{L2 Functions} \\
\midrule
Scroll & $(x_1, y_1), (x_2, y_2)$ & Navigate \\
Primary click & $(x, y)$ & Select \\
Double primary click & $(x, y)$ &Select \\
Secondary click & $(x, y)$ &Select \\
Point & $(x_1, y_1), (x_2, y_2)$ & Navigate\\
Drag drop & $(x_1, y_1), (x_2, y_2)$ & Drag\\
\midrule
Shortcut & Modifier + Non-Modifier & Type, Navigate, Select, Program  \\
& Single Key & Program  \\
Keypress & Key & Type, Navigate, Select \\
\bottomrule
\end{tabular}
\caption{Mapping from L1 Operations and specific details to their corresponding L2 Functions. Mouse operations map to fixed functions, while keyboard operations may map to different functions depending on the executed key or shortcut.}
\Description{Table mapping L1 Operations and their specific input details to L2 Functions. Mouse operations map deterministically: scrolling and pointing map to Navigate, primary, double-primary, and secondary clicks map to Select, and drag-and-drop maps to Drag. Keyboard operations are context-dependent: shortcuts can map to Type, Navigate, Select, or Program, while keypresses can map to Type, Navigate, or Select. The table also shows the low-level details retained for each operation, such as cursor coordinates for mouse actions and executed keys or modifiers for keyboard actions.}
\label{tab:operation-function-mapping}
\end{table}

\subsection{Methods (L3)}
L3 captures the higher-level procedure a sequence of interactions carries out, allowing traces with different low-level executions to be compared as the same method. It groups contiguous L2 interactions that contribute to one immediate objective into a \textit{Method}, together with the task-relevant \textit{Specific Entity} and the \textit{UI Container} in which it occurs, denoted as \texttt{Method} $\langle$Specific Entity$\rangle$ in $\langle$UI Container$\rangle$. For example, one user may navigate through several menus to add page numbers while another uses a keyboard shortcut - their L2 sequences differ, but both are represented at L3 as \texttt{add} $\langle$page numbers$\rangle$ in $\langle$text document$\rangle$. L3 therefore makes the shared procedure comparable while L1 and L2 retain the differences in how it was executed and the intermediate outcomes observed respectively.

L3's \textit{Method} describes what procedure is being performed, \textit{Specific Entity} identifies what it is being performed on, and \textit{UI Container} captures the broader interface context. Methods and UI Containers use fixed vocabularies so equivalent procedures and contexts receive comparable labels across users, tasks, and datasets. Specific Entities remain open-vocabulary because they depend on the particular task, such as page numbers, homework.docx, or a named product. This separation keeps the fixed labels general: the same Method can apply to different task-specific entities, while UI Containers describe types of interfaces rather than particular applications. For example, \texttt{add} can describe adding different entities within a \texttt{text document}, without introducing a new Method for each entity.

We construct the Method vocabulary by open-coding a subset of interaction traces from a computer use dataset~\cite{mohanbabu2026a11y}. We first assigned descriptive labels to sampled interaction segments, then consolidated labels that described the same procedure and generalized labels that were specific to an application or task. This resulted in a fixed vocabulary of 14 Methods (refer to Appendix ~\ref{sec:methods}, ~\ref{sec:ui-containers} for L3 Methods and UI Containers vocabulary).
The traces used for developing this vocabulary are not included in the evaluation set. We similarly define 16 UI Containers spanning applications, websites, documents, windows, and other content areas. For example, labels such as \texttt{text document} describe a type of working context rather than a particular program or file, allowing the same container label to be reused across datasets. Both vocabularies include \texttt{Unknown} when the interaction trace does not provide enough evidence to assign a predefined label confidently.

\subsection{Goals (L4)}
Goals provide the highest level of abstraction in \system{}, grouping contiguous L3 Methods that jointly accomplish the same task-level objective. This corresponds to the level of semantic abstraction commonly inferred from interaction traces in prior work~\cite{zhang2025summact,jiang2026inducing}. Because Goals are task-dependent, they use an open vocabulary to accommodate diverse objectives across datasets.

\begin{table}[]
\centering
\small
\begin{tabular}{llll}
\toprule
\textbf{L\#} & \textbf{Label} & \textbf{Input} & \textbf{Inference} \\
\midrule
L1 & Operations & IT  & Rule-based\\
L2  & Functions & IT  & Rule-based\\
 & Direct UI Targets & IT + Targets' list & LLM\\
& Consequences & IT + Consequences' list & LLM\\
& Indirect UI Targets & IT + Targets' list & LLM\\
L3 & Methods & IT + Methods' list & LLM\\
& Specific UI Entity & IT & LLM\\
 & UI Containers & IT + UI Containers' list & LLM\\
\bottomrule
\end{tabular}
\caption{Overview of how \system{} labels are inferred for actions in an interaction trace (IT). For each taxonomy level, we show the information provided as input and whether labels are assigned using deterministic rules or LLM inference from a fixed candidate label set.}
\Description{Table summarizing how MAP assigns labels at each abstraction level. L1 Operations and L2 Functions are inferred from the interaction trace using deterministic rules. The remaining L2 fields—Direct UI Targets, Consequences, and Indirect UI Targets—are inferred by an LLM using the interaction trace plus the corresponding fixed candidate label sets. At L3, Methods and UI Containers are likewise inferred from fixed vocabularies, while UI Entity is inferred from the interaction trace without a fixed candidate list.}
\label{tab:map-inference-source}
\end{table}

\section{\system{} Similarity Metrics}

The \system{} taxonomy provides a common semantic representation of interaction traces, but comparing users also requires specifying \textit{what aspect} of their behavior is similar. We capture three complementary dimensions: \textbf{sequence similarity} measures whether two traces progress through a task in a similar order, \textbf{pattern similarity} measures whether they exhibit similar recurring local behaviors, and \textbf{effort similarity} measures whether they allocate comparable time and actions to the same parts of a workflow. After annotating two traces with \system{}, we compute each metric independently at L1--L3, allowing similarity to be examined both by behavioral dimension and level of abstraction.

\subsection{\system{}-Sequence Similarity}

\system{}-Sequence similarity captures whether two traces perform interactions in a similar \textit{order}, distinguishing traces that contain similar interactions but arrange them differently. Prior work has used edit-distance-based methods to compare text~\cite{lcvenshtcin1966binary} and interaction sequences in software tutorials~\cite{kong2012delta,pavel2013browsing}. We similarly use Needleman--Wunsch global alignment~\cite{needleman1970general} to identify which interactions correspond across the two complete traces and where they differ. At each level, identical labels receive a match score of $+1$, while mismatches and gaps receive a penalty of $-1$. We define sequence similarity as the proportion of positions in the resulting alignment whose labels match exactly. The alignment additionally identifies interactions that are substituted, added, or omitted between traces.

\subsection{\system{}-Pattern Similarity} 
\system{}-Pattern similarity measures whether two \system{}-labeled traces exhibit similar \emph{local interaction patterns}, even when their overall task paths differ  (\textit{e.g.}, two traces that use hotkey shortcuts vs. point-and-click). This complements sequence similarity by capturing recurring behavioral tendencies. For each of \system{} levels from L1-L3, we extract n-grams of lengths 1, 2, and 3 and compute cosine similarity between n-gram frequency vectors at each length, similar to bag-of-words metrics frequently used for text comparison~\cite{harris1954distributional}.
We then compute a weighted average that rewards longer n-grams to capture more specific behavioral styles between the two traces. This metric can reveal similarity between user interaction traces within task and across tasks. \\ 

\subsection{\system{}-Effort Similarity}

\system{}-Effort similarity measures whether two users allocate comparable interaction effort to equivalent parts of a task: two traces may be semantically identical yet differ substantially in cost -- \textit{e.g.}, formatting a document in 5 actions and 10 seconds versus 15 actions and 40 seconds. We characterize effort by action count and time, two standard measures of interaction efficiency in HCI~\cite{card1980keystroke}, and compute similarity independently at each \system{} level $l$.

When computing the similarity between traces $A$ and $B$, let $S_l$ be the union of semantic labels appearing in either trace at level $l$ (\textit{e.g.}, an \texttt{operation} at L1 or \texttt{method} $\langle$UI container$\rangle$ at L3), and let $a^{A}_{l,s}$ and $t^{A}_{l,s}$ (and likewise for $B$) denote the number of actions and total time associated with label $s$. For each label, we compute the squared log-ratios
$d^{a}_{l,s}=\ln^{2}\big((a^{A}_{l,s}+1)/(a^{B}_{l,s}+1)\big)$ and
$d^{t}_{l,s}=\ln^{2}\big((t^{A}_{l,s}+1)/(t^{B}_{l,s}+1)\big)$,
which capture proportional rather than absolute differences~\cite{tornqvist1985should} and are symmetric in $A$ and $B$. We combine the action and time differences for each label as
$\mathrm{effort}_{l,s}=(d^{a}_{l,s}+1)(d^{t}_{l,s}+1)$,
and sum across labels to obtain
$\mathrm{effortDiff}_{l}(A,B)=\sum_{s\in S_l}\mathrm{effort}_{l,s}$.
We then map this dissimilarity to
$\mathrm{effortSim}_{l}(A,B)=100\,e^{-\mathrm{effortDiff}_{l}(A,B)/100}$.
A difference in actions, time, or both lowers the score, and computing it at each level distinguishes traces that allocate effort differently across semantic labels.

\section{\system{} Pipeline}
The \system{} pipeline builds the hierarchy from low-level Operations (L1) to higher-level Goals (L4). It assigns labels deterministically when they can be derived directly from recorded inputs, and uses an LLM to infer labels that require interaction context.

\subsection{\system{} Annotation}

\system{} annotates traces hierarchically from L1 to L4, with each level using the recorded interaction trace and labels from preceding levels. Before annotation, dataset-specific action names are mapped to the common L1 Operations defined in Sec.~\ref{sec:taxonomy}. Both L1 Operations and L2 Functions labels are assigned deterministically. The remaining L2 fields (Direct UI Target, Consequence, and Indirect UI Target) are selected from their fixed vocabularies using the surrounding interaction context (refer to Appendix ~\ref{sec:prompts} for L2, L3, L4 prompts).

Annotation uses information available in the trace, including the active application or window, interacted or focused UI element, and preceding actions and labels. Screen-reader state is included when available and needed to interpret keyboard interaction. We exclude task descriptions, task outcomes, user metadata, and screenshots, allowing the same procedure to be applied across datasets with different recording setups. This differs from approaches that additionally use screenshots for segmentation or goal inference~\cite{wang2025ai}.

At L3, \system{} divides the trace into \textit{segments}: contiguous runs of interactions that carry out the same Method on the same Specific Entity within the same UI Container. The model jointly determines these boundaries and assigns the three L3 fields. A new segment begins when the Method, Specific Entity, or UI Container changes, or when the active interface changes. The full segmentation rules and L3 prompt are provided in Appendix~\ref{sec:prompts}. At L4, contiguous L3 segments that contribute to the same task-level objective are grouped under a Goal.
We annotate the non-deterministic semantic labels from L2--L4 using Claude Sonnet 5. Refer to supplementary materials for all prompts and full \system{} vocabulary.

\newcommand{\msd}[2]{#1 {\scriptsize$\pm$ #2}}

\begin{table*}[t]
\centering
\small
\setlength{\tabcolsep}{4pt}

\begin{tabular}{@{}llrrcccccc@{}}
\toprule
& & & &
\multicolumn{4}{c}{\textbf{L2}} &
\multicolumn{2}{c}{\textbf{L3}} \\
\cmidrule(lr){5-8}
\cmidrule(lr){9-10}

\textbf{Dataset} &
\textbf{Group} &
\textbf{Traces} &
\shortstack{\textbf{Actions}\\\textbf{/Trace}} &
\textbf{Function}$^{*}$ &
\shortstack{\textbf{Direct UI}\\\textbf{Target}} &
\textbf{Consequence} &
\shortstack{\textbf{Indirect UI}\\\textbf{Target}} &
\textbf{Method} &
\shortstack{\textbf{UI}\\\textbf{Container}} \\

\midrule

A11y-CUA
& SU
& 10 (534)
& \msd{53.4}{29.4}
& 100.0
& \msd{79.7}{15.2}
& \msd{90.0}{6.7}
& \msd{73.5}{15.9}
& \msd{82.6}{22.4}
& \msd{93.0}{7.5} \\

& BLVU
& 5 (324)
& \msd{64.8}{20.1}
& 100.0
& \msd{70.2}{16.6}
& \msd{86.9}{12.1}
& \msd{79.4}{11.6}
& \msd{81.6}{29.3}
& \msd{98.7}{2.0} \\

& CUA-D
& 5 (308)
& \msd{61.6}{63.3}
& 100.0
& \msd{86.2}{4.8}
& \msd{79.2}{14.9}
& \msd{82.8}{7.4}
& \msd{93.0}{10.1}
& \msd{79.6}{29.4} \\

\addlinespace

AgentSynth
& ---
& 5 (978)
& \msd{195.6}{95.2}
& 100.0
& \msd{91.0}{7.9}
& \msd{88.4}{15.6}
& \msd{85.0}{22.4}
& \msd{63.4}{39.8}
& \msd{80.3}{24.2} \\

AgentNet
& ---
& 5 (198)
& \msd{39.6}{31.1}
& 100.0
& \msd{84.2}{19.4}
& \msd{89.0}{7.2}
& \msd{69.6}{30.3}
& \msd{73.9}{41.7}
& \msd{75.4}{42.2} \\

\midrule

\multicolumn{2}{@{}l}{\textbf{Average}}
&
&
\msd{78.1}{72.0}
& 100.0
& \msd{81.8}{14.7}
& \msd{87.3}{11.0}
& \msd{77.3}{18.2}
& \msd{79.5}{28.8}
& \msd{86.7}{23.2} \\

\bottomrule
\end{tabular}

\caption{
Agreement between \system{} and manual annotations across 30 traces and 2,342 actions.
Agreement values report mean percentage agreement across traces ($\pm$ SD).
\textbf{Traces} reports the number of traces, with total actions in parentheses, and \textbf{Actions / Trace} reports mean actions per trace ($\pm$ SD).
$^{*}$Function, like Operation (L1), is assigned deterministically and is therefore 100\% by construction.
A11y-CUA groups complete the same tasks, whereas AgentSynth and AgentNet contain traces from different tasks and applications and are included to evaluate generalization.}

\Description{Table reporting agreement between MAP and manual annotations across datasets and user groups. Rows include sighted users, blind/low-vision users, and CUA-D traces from A11y-CUA, plus AgentSynth and AgentNet. Columns report traces and actions per trace, followed by L2 agreement for Function, Direct UI Target, Consequence, and Indirect UI Target, and L3 agreement for Method and UI Container. Function is 100\% by construction because it is assigned deterministically.}
\label{tab:agreement}
\end{table*}

\subsection{Pipeline Evaluation}
We evaluate how accurately \system{} annotates interaction traces and how well it generalizes across users, agents, tasks, and applications.

\subsubsection{Datasets} 
We evaluate \system{} on three computer use datasets: A11y-CUA~\cite{mohanbabu2026a11y}, AgentSynth~\cite{xie2026agentsynth}, and AgentNet~\cite{yang2026agentnet}. A11y-CUA contains the same 60 tasks completed by 16 participants (8 sighted and 8 blind users) and by agents under a default condition (CUA-D) or screen reader condition (CUA-SR). Its timestamps and detailed interaction context support both annotation evaluation and the similarity analyses reported later.
We use a subset of A11y-CUA to construct the L3 Method vocabulary. However, these traces are disjoint from the A11y-CUA traces used in our annotation evaluation. We additionally evaluate on AgentSynth and AgentNet to test whether \system{}'s labels apply beyond the dataset used during vocabulary development. AgentSynth contains synthetic agent traces, while AgentNet contains human demonstrations; together they introduce applications and interaction patterns beyond those in A11y-CUA. Because neither provides the timing information needed for time-based effort similarity, we use all three datasets to evaluate annotation accuracy and vocabulary coverage, and A11y-CUA for the cross-user and agent similarity analyses.

\subsubsection{Manual Annotation}

We manually annotated 30 traces containing 2,342 actions: 20 from A11y-CUA (10 SU, 5 BLVU, and 5 Agent-Default), 5 from AgentSynth, and 5 from AgentNet. A11y-CUA traces were sampled randomly within each condition. For AgentSynth and AgentNet, we selected traces from five different applications in each dataset to evaluate \system{} beyond the applications represented in A11y-CUA.

For each action, we annotated L2's \texttt{Direct UI Target}, \texttt{Consequence}, and \texttt{Indirect UI Target}. We also divided each trace into L3 segments and annotated each segment's \texttt{Method}, \texttt{Specific Entity}, and \texttt{UI Container}. We did not manually annotate L1 \texttt{Operation} or L2 \texttt{Function}, which are assigned deterministically. For A11y-CUA, annotators inspected the action trace together with the recorded video and audio, including screen-reader output when present. AgentSynth and AgentNet do not provide corresponding videos, so annotators instead used the interaction trace together with the textual observations and reasoning recorded with the actions.

Annotators followed a codebook specifying how to apply the L2 and L3 labels, including label definitions, and L3 segmentation rules (full codebook is provided in the Appendix ~\ref{sec:vocabulary}). Two researchers independently annotated three A11y-CUA pilot traces (249 actions; 1,245 labels across the five closed-vocabulary fields). We refined the annotation instructions and examples based on disagreements until reaching Cohen's $\kappa=0.79$, conventionally interpreted as substantial agreement~\cite{landis1977measurement}. One researcher then annotated the remaining 27 traces using the finalized codebook. Annotation took approximately 25 minutes per trace, or about 12 hours in total.

We then ran the complete \system{} pipeline on the same 30 traces without access to the manual annotations. We compared its L2 and L3 annotations with the manual annotations to evaluate semantic labeling and L3 segmentation.

\subsubsection{Results}

\noindent\textbf{Accuracy.}
\system{} achieves generally high exact-match accuracy against manual annotations (Table~\ref{tab:agreement}). Across the 28 traces, mean per-trace accuracy is 81.8\% ($\sigma=14.7$) for \texttt{Direct UI Target}, 87.3\% ($\sigma=11.0$) for \texttt{Consequence}, and 77.3\% ($\sigma=18.2$) for \texttt{Indirect UI Target}. At L3, accuracy is 79.5\% ($\sigma=28.8$) for \texttt{Method} and 86.7\% ($\sigma=23.2$) for \texttt{UI Container}.

L3 scores are particularly sensitive to segmentation: when \system{} and the annotator place a segment boundary differently, the Method or UI Container assigned to several neighboring actions can all count as mismatches even when they identify the same broader procedure. Across 136 inspected L3 errors, 38.2\% are boundary shifts, 36.8\% merge adjacent manual segments, and 18.4\% split one manual segment; only 6.6\% have matching boundaries but a different Method label. Thus, 93.4\% of the inspected L3 errors concern how interactions are grouped rather than which Method describes an identically bounded segment. For example, a sequence that opens one browser, dismisses it, switches browsers, and then searches can reasonably be divided into several procedures, while \system{} groups the sequence into one \texttt{search} segment. These cases reflect differences in segmentation granularity more often than confusion between unrelated Methods.

L2 errors also do not usually determine the L3 result. L3 errors occur at nearly the same rate when the L2 tuple is incorrect (36.2\%) as when it is correct (37.3\%): 393 actions have an incorrect L2 annotation but a correct L3 annotation, while 642 have a correct L2 annotation but an incorrect L3 annotation. This indicates that L3 annotation is driven primarily by the surrounding sequence and where procedure boundaries are placed, rather than by any single L2 label. \\

\noindent\textbf{Coverage.}
We separately examine whether the taxonomy contains labels applicable to the behaviors observed in each dataset, independent of whether \system{} selects the correct label. Across A11y-CUA's 540 traces and 60,974 actions, 99.2\% of closed-vocabulary fields can be assigned a predefined label rather than \texttt{Unknown}, and 118 of the 126 taxonomy labels appear at least once (93.7\%). The same predefined vocabularies cover 99.9\% of fields in AgentSynth and 99.8\% in AgentNet, indicating that behaviors in these datasets rarely require labels outside the taxonomy. AgentSynth and AgentNet instantiate smaller subsets of the vocabulary - 70/126 (55.6\%) and 71/126 (56.3\%), respectively - because they contain a narrower range of observed interaction behaviors than A11y-CUA. Across the three datasets, 120 of the 126 taxonomy labels appear at least once (95.2\%).

\section{Results}

\begin{table*}[t]
\centering
\small
\setlength{\tabcolsep}{0.4em}
\renewcommand{\arraystretch}{1.18}

\label{tab:results-summary}

\begin{tabularx}{\textwidth}{
    @{}
    >{\centering\arraybackslash}p{0.045\textwidth}
    >{\centering\arraybackslash}p{0.055\textwidth}
    >{\raggedright\arraybackslash}p{0.30\textwidth}
    >{\centering\arraybackslash}p{0.09\textwidth}
    >{\raggedright\arraybackslash}X
    @{}
}
\toprule
\textbf{RQ} &
\textbf{Level} &
\textbf{Prior work finding} &
\textbf{Confirmed by \system{}} &
\textbf{New finding from \system{}} \\
\midrule

RQ1 & L1 &
SUs primarily use mouse input, whereas BLVUs use keyboard input~\cite{bigham2007webinsitu,mohanbabu2026a11y}. &
Yes &
BLVUs repeat similar keyboard-action patterns more consistently, but vary more in the overall order of their actions across tasks. \\

RQ1 & L2 &
BLVUs frequently use sequential and exploratory navigation and use multiple navigation strategies~\cite{takagi2007analysis,bigham2007webinsitu,borodin2010more}. &
Yes &
BLVUs require substantially more effort for navigation, while typing and selection are similar across groups. \\

RQ1 & L3 &
BLVUs and SUs differ in how they formulate and reformulate search queries~\cite{sahib2012comparative}. &
Yes &
BLVUs have longer search queries but SUs query more often per task. Additionally, despite these lower-level differences, SUs and BLVUs usually perform the same Methods to complete a task. \\

RQ1 & L4 &
BLVUs adopt coping and recovery strategies when encountering accessibility barriers~\cite{vigo2013coping,bigham2007webinsitu}. &
Yes &
BLVUs more often add help-seeking goals, whereas SUs more often add verification and window-management goals. \\

\midrule

RQ2 & L1 &
Agents differ from humans in their low-level execution, and changing an agent's access mode changes its interaction behavior~\cite{wang2025ai,mohanbabu2026a11y}. &
Yes &
Matching access mode does not guarantee human-like execution: CUA-SR uses far fewer distinct keyboard inputs than BLVUs, while CUA-D remains less similar to SUs. \\

RQ2 & L2 &
Screen reader use increases agent actions, and agents often type long inputs where humans navigate through the interface~\cite{mohanbabu2026a11y}. &
Yes &
Humans interleave typing with navigation and screen reader management more often, whereas agents more often type uninterruptedly and then select UI targets. \\

RQ2 & L3 &
Humans and agents can follow similar high-level task steps despite different low-level execution~\cite{wang2025ai}. &
Yes &
CUA-D \texttt{verifies} more often than SUs, whereas SUs \texttt{move} and \texttt{dismiss} interface elements more often. Both agents perform Methods such as \texttt{open} and \texttt{edit} at similar rates to their corresponding human groups.\\

RQ2 & L4 &
Humans and agents can pursue similar task goals despite differences in how they execute them~\cite{wang2025ai}. &
Yes &
Agents usually reach the same primary task goal as humans but omit additional goals such as help-seeking, or closing extra windows and tabs. \\

\bottomrule
\end{tabularx}

\caption{Summary of prior findings recovered by \system{} and new findings revealed by its multi-level representation.}
\Description{Table summarizing how \system{} reproduces findings from prior work and reveals additional findings across four abstraction levels for two research questions. Columns list the research question, \system{} level, corresponding prior finding, whether \system{} confirms it, and a new finding enabled by the multi-level representation. For RQ1, comparing sighted users (SUs) and blind/low-vision users (BLVUs), L1 confirms mouse- versus keyboard-dominant input and shows more repeated keyboard patterns among BLVUs; L2 confirms heavier navigation by BLVUs and localizes their additional effort to navigation rather than typing or selection; L3 confirms differences in search behavior while showing that both groups often use the same Methods; and L4 confirms coping and recovery behavior while revealing different auxiliary Goals such as help-seeking, verification, and window management. For RQ2, comparing humans and computer-use agents, L1 shows that matching access mode does not yield human-like low-level execution; L2 shows different patterns of typing, navigation, and selection; L3 shows convergence on similar Methods despite execution differences; and L4 shows that agents usually reach the same primary Goals but omit auxiliary human Goals such as help-seeking and closing extra windows or tabs.}
\end{table*}

We demonstrate \system{} by investigating two research questions from prior work. \textbf{RQ1: How do sighted and blind users differ in how they complete tasks with computers?}~\cite{takagi2007analysis,bigham2007webinsitu,sahib2012comparative,vigo2013coping,mohanbabu2026a11y} Comparing these groups can characterize accessibility gaps in computer interaction and identify where interfaces create additional effort for blind users. \textbf{RQ2: How closely do computer use agent workflows align with human workflows?}~\cite{wang2025ai,mohanbabu2026a11y} This question is increasingly relevant as agents are used to simulate users for applications such as usability testing and design evaluation~\cite{holter2026uxcascade,rieder2026simab}. Comparing agent and human traces can therefore show which aspects of human behavior agents reproduce and which they omit.

We investigate these questions using A11y-CUA~\cite{mohanbabu2026a11y}, analyzing 30 tasks performed by 16 users (SU1--SU8, BU1--BU8) and two agents (CUA-D and CUA-SR). We compare within- and across-group behavior using \system{}-Sequence, Pattern, and Effort similarity at L1--L4. We report pairwise similarity matrices and cluster participants and agents using K-means over the three similarity metrics at each level. For each clustering, we report the Silhouette score, where higher values indicate that traces are more similar within clusters and more distinct across clusters, to quantify how strongly each abstraction level separates behavioral groups.

\subsection{How do blind and sighted users differ in how they complete tasks with computers?}

\noindent{\textbf{Operations (L1): Blind and sighted users differ in how they interact with the interface.}} \
Prior work has shown that SUs and BLVUs differ in their input methods: sighted users primarily use mouse actions, whereas blind screen reader users use keyboard actions~\cite{bigham2007webinsitu,mohanbabu2026a11y}. \system{} confirms this at L1. The two groups are separated across all three similarity metrics, especially for Effort and Pattern (Silhouette: Sequence=0.40, Effort=0.70, Pattern=0.77). SUs use \texttt{point} and \texttt{primary click} for 32.9\% and 25.1\% of their actions, respectively, while BLVUs interact entirely through keypresses and shortcuts. However, SUs are not only mouse users: keypresses still account for 28.0\% of their actions, showing that they regularly combine mouse and keyboard input.

The similarity metrics further show that BLVUs use more similar recurring action patterns, but vary more in their overall action sequences. Their most common bigrams are \texttt{keypress}$\rightarrow$\texttt{keypress} (70.9\%) and \texttt{keypress}$\rightarrow$\texttt{shortcut} (16.3\%). SU traces, in contrast, reflect their use of both input methods: their most common bigrams include \texttt{keypress}$\rightarrow$\texttt{keypress} (24.7\%) and \texttt{point}$\rightarrow$\texttt{primary click} (22.4\%). BLVUs therefore have higher within-group Pattern similarity than SUs (93.8 vs.\ 86.2), but lower Sequence similarity (46.6 vs.\ 52.9). Overall, L1 mainly captures differences in how the two groups interact with the interface and at L2, we examine what those actions are doing.\\

\noindent{\textbf{Functions \& Consequences (L2): Accessibility differences are concentrated in navigation.}} 
Prior observational studies show that screen reader users frequently use sequential and exploratory navigation to move through and find content, often using scanning and multiple keyboard navigation strategies~\cite{takagi2007analysis,bigham2007webinsitu,borodin2010more}. \system{} confirms this difference at L2 and localizes where it occurs. \texttt{navigate} accounts for 70.8\% of BLVU actions, compared with 39.7\% of SU actions. The difference is even clearer in Function$\rightarrow$Consequence pairs: \texttt{navigate}$\rightarrow$\texttt{focus}, where navigation moves focus to another interface element, accounts for 47.8\% of BLVU actions but only 1.1\% of SU actions. \texttt{navigate}$\rightarrow$\texttt{none}, where a navigation command probes the interface but produces no recorded interface consequence, accounts for another 14.6\% of BLVU actions and is nearly absent for SUs. The difference in effort is also concentrated in navigation: BLVUs use 9.72 actions on average per navigation instance compared with 1.30 for SUs, while differences for \texttt{select} (1.22 vs.\ 1.15) and \texttt{type} (5.80 vs.\ 6.26) are much smaller.

\begin{figure*}
    \centering
    \includegraphics[width=6in]{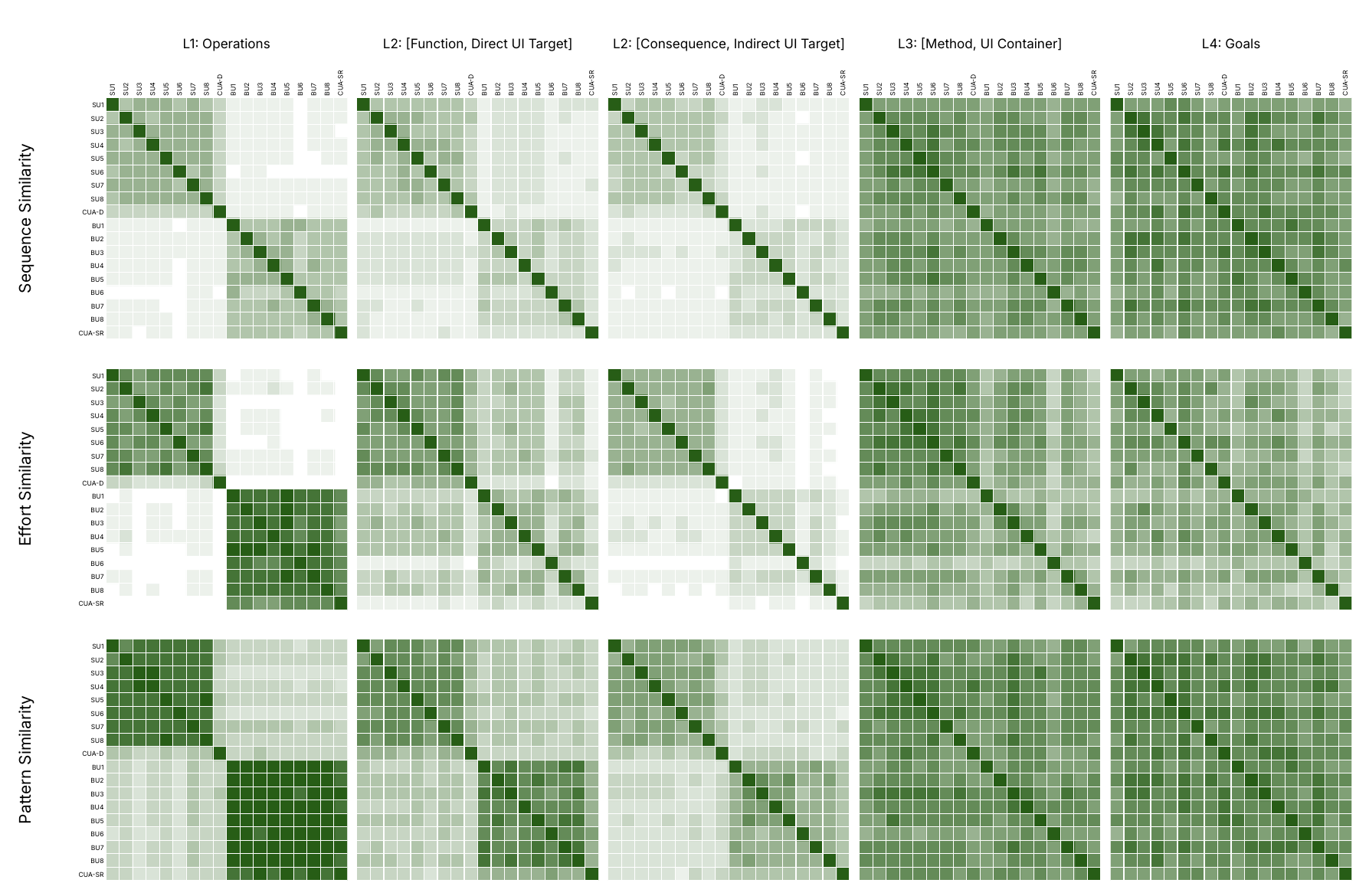}
    \caption{
    Pairwise Sequence, Effort, and Pattern Similarity among sighted users (SU), blind users (BU), CUA-D, and CUA-SR across \system{} levels. Darker cells indicate higher similarity. Strong within-group blocks at L1 and L2 weaken at L3 and L4, showing that SU and BU differ more in low-level execution than in higher-level methods and goals. CUA-D and CUA-SR generally resemble SU and BU, respectively, at lower levels.
    }
    \Description{A 3-by-5 grid of pairwise similarity heatmaps. Rows show Sequence, Effort, and Pattern Similarity; columns progress from L1 Operations through two L2 representations, L3 Methods and UI Containers, and L4 Goals. Each heatmap compares eight sighted users, CUA-D, eight blind users, and CUA-SR, with darker green indicating greater similarity. At L1, and to a lesser extent L2, dark blocks appear within the sighted and blind groups while cells between groups are much lighter. These blocks become substantially less distinct at L3 and L4, where similarity is more uniform across users. CUA-D tends to resemble sighted users and CUA-SR blind users at the lower levels.}
    \label{fig:sim-mat}
\end{figure*}

Looking across L1 and L2 using the \system{} taxonomy, shows that the two groups can \textit{perform the same function using very different operations}. For example, when navigating to a \ui{list item}, SUs use \texttt{point} in 99\% of cases, whereas BLVUs use \texttt{keypress} (89\%) or \texttt{shortcut} (11\%). The reverse is also true: \textit{the same operation can serve different functions}. Among SUs, 81.9\% of keypresses are used to \texttt{type}, whereas among BLVUs, 75.6\% are used to \texttt{navigate}. L2 therefore reveals differences that are hidden at L1. Although BLVUs all use only keyboard input, their within-group Effort similarity falls from 86.5 at L1 to 46.5 at L2 (Fig.~\ref{fig:sim-mat}), showing greater variation in the effort required to carry out functions such as navigation. Thus, the main difference at L2 is not simply that BLVUs use the keyboard, but that much more of their interaction is devoted to navigating and searching through the interface. \\

\noindent{\textbf{Methods (L3): Blind and sighted users converge on Methods, but effort differences remain.}} \
Prior work has shown that blind and sighted users can differ in how they search, including how they formulate and reformulate queries~\cite{sahib2012comparative}. We observe similar differences: BLVU query bursts contain 17.3 characters on average compared with 9.9 for SUs, while SUs issue slightly more query bursts per task (2.5 vs.\ 2.2).

However, across the full set of tasks, the two groups become much more similar at L3. Across-group Sequence and Pattern similarity reach 68.8 and 72.5, close to the corresponding within-group similarities (SU: 72.0/76.8; BLVU: 67.5/71.4) (Fig.~\ref{fig:sim-mat}). In 23 of 30 tasks, similarity between SUs and BLVUs increases from L1 to L3 by 42.3 points on average ($\sigma=22.1$). Thus, actions that look very different at L1 often correspond to the same Methods at L3.

The groups can still differ substantially in how those Methods are carried out. For example, for the Method \texttt{open}, SU actions consist of 50\% \texttt{select} and 40\% \texttt{navigate} at L2, whereas BLVU actions consist of 9\% \texttt{select} and 73\% \texttt{navigate}. Effort also varies more among BLVUs: within-BLVU Effort similarity is 53.6, compared with 78.1 for SUs (Table~\ref{tab:similarity}). This shows that even when BLVUs perform similar Methods, they differ more from one another in how many lower-level actions those Methods require.

The groups do not always become more similar at L3. In 7 of 30 tasks, SU--BLVU similarity decreases from L1 to L3 because the groups perform different Methods even when their lower-level actions look similar. In one task, for example, similarity decreases from 82.3 at L1 to 42.1 at L3 because SUs additionally perform \texttt{dismiss} and \texttt{verify}, Methods that are absent from the BLVU traces. These differences are difficult to see at L1, where the underlying clicks and keypresses may still look similar. L3 therefore shows both where different interactions accomplish the same Methods and where the groups actually follow different Methods.\\

\noindent \textbf{Goals (L4): User groups converge on task goals, with differences in auxiliary goals.} \\
Prior work has shown that blind users adopt coping strategies when they encounter accessibility barriers~\cite{vigo2013coping,bigham2007webinsitu}. We observe evidence of these strategies at L4. In particular, BLVUs take on additional goals to seek help from online resources while completing tasks: 9 of the 22 goals observed among BLVUs involve help-seeking, including 5 that use a screen reader assistant. SUs also take on additional goals, but of a different kind: 8 of their 20 observed goals involve verification or window management. Thus, the differences between the groups at L4 are concentrated in these additional goals rather than in the main goals needed to complete the task.
Otherwise, SU and BLVU traces largely converge at L4. Across-group Sequence and Pattern similarity reach 70.7 and 74.7, and both groups average about 1.6 goals per trace. Thus, despite differences in operations, navigation, and Methods at lower levels, the two groups generally work toward the same goals for the same tasks (Fig.~\ref{fig:sim-mat}). Differences at L4 instead reflect additional goals along the way.\\

\subsection{How closely do computer use agent workflows align with human workflows?}

Prior work shows that humans and agents can follow similar high-level steps despite differences in how they execute them~\cite{wang2025ai}. A11y-CUA further shows that changing an agent's access mode substantially changes its behavior and performance~\cite{mohanbabu2026a11y}. We use \system{} to connect these findings by examining \emph{where} agents align with humans across abstraction levels. We compare CUA-D with SUs and CUA-SR with BLVUs, using the corresponding within-human similarity as a baseline. \\

\noindent\textbf{Operations (L1): Agents mirror human access modality, but not execution.} 
Prior A11y-CUA analysis showed that neither agent condition fully reproduces the interaction of its corresponding human group. CUA-D uses mouse and keyboard input like SUs but differs in its mix of actions, while CUA-SR uses keyboard input like BLVUs but differs in the specific keys and shortcuts it uses~\cite{mohanbabu2026a11y}. \system{} shows how these differences appear when human and agent traces are represented using the same L1 operations.
CUA-D is much less similar to SUs than SUs are to one another: its Pattern similarity with SUs is 42.8, compared with 86.2 within SUs. Although both use mouse and keyboard input, their recurring operation patterns differ. For example, CUA-D uses \texttt{double primary click} more often than SUs (8.8\% vs.\ 4.5\%) and does not use \texttt{drag drop} at all (vs.\ 1.5\% for SUs).

CUA-SR, in contrast, is much more similar to BLVUs at L1. Its Pattern similarity with BLVUs is 88.7, reaching 95\% of the 93.8 similarity observed between BLVUs themselves. This high similarity reflects their shared reliance on \texttt{keypress} and \texttt{shortcut} operations. However, the specific keyboard inputs still differ: CUA-SR uses 102 distinct keystrokes and shortcuts, compared with 225 for BLVUs. Thus, L1 captures the shared access mode between agents and their corresponding human groups, while differences in their specific execution remain within those operations.\\

\noindent\textbf{Functions \& Consequences (L2): Screen reader use shows similar increases in actions for agents and humans. Agents make fewer corrections than humans.} 
A11y-CUA found that screen reader use substantially increases agent actions~\cite{mohanbabu2026a11y}: CUA-SR performs 3.0$\times$ as many actions as CUA-D, closely matching the 2.9$\times$ increase from SUs to BLVUs. \system{} shows where this increase occurs at L2. For agents, actions with no recorded interface consequence increase from 5.7\% for CUA-D to 16.7\% for CUA-SR; for humans, they similarly increase from 4.3\% for SUs to 21.1\% for BLVUs. Thus, much of the additional interaction introduced by screen reader use occurs in actions that do not immediately produce a recorded interface change.

A11y-CUA also observed that agents often type long URLs or commands to navigate or complete tasks, whereas humans more often navigate through the interface~\cite{mohanbabu2026a11y}. \system{} makes this difference visible at L2. For CUA-D, \texttt{type}\ui{text area}$\rightarrow$\texttt{text input}\ui{text area} accounts for 27.1\% of its L2 patterns, compared with 10.4\% for SUs, showing that the agent spends more of its interaction in uninterrupted typing. When typing is interrupted, 86\% of CUA-D interruptions are followed by \texttt{select} and only 2\% by \texttt{navigate}, compared with 44\% and 47\% for SUs. CUA-SR also almost never produces the \texttt{mute} consequence that BLVUs use to manage screen reader speech (0.02 vs.\ 3.26 per 100 actions). Thus, similar increases in total actions do not imply similar interaction behavior: humans more often interleave typing with navigation and screen reader management, whereas agents rely more on extended typing and direct selection. This distinction matters when agents are used to simulate users for interface evaluation~\cite{holter2026uxcascade,rieder2026simab}: matching task completion or action count does not necessarily mean reproducing how humans interact with the interface.\\

\noindent\textbf{Methods (L3): Humans and agents converge on Methods despite different low-level execution.} \
Prior work finds that humans and agents can complete tasks using similar high-level steps even when their low-level actions differ~\cite{wang2025ai}. \system{} confirms this at L3. SU--CUA-D Pattern similarity rises from 42.8 at L1 to 73.5 at L3, reaching 96\% of the within-SU similarity (76.8). BLVU--CUA-SR similarity shows the same pattern, reaching 70.7 at L3, or 99\% of the within-BLVU similarity (71.4). In other words, by L3, each agent is almost as similar to its corresponding human group as humans within that group are to one another.

The L3 labels show where this similarity comes from. For CUA-D and SUs, \texttt{open} accounts for 36.8\% and 38.4\% of L3 Method instances, respectively, while \texttt{edit} accounts for 18.4\% and 17.7\%. Thus, although humans and agents differ substantially in the lower-level actions used to carry out tasks, they often perform the same L3 Methods at similar rates.\\

\noindent\textbf{Goals (L4): Agents usually reach the same primary task goals as humans, but omit auxiliary goals.} \
Prior work finds that humans and agents can pursue similar high-level task goals despite differences in lower-level execution~\cite{wang2025ai}. \system{} confirms this most clearly for CUA-D. Its Pattern similarity with SUs is 77.5, comparable to the within-SU similarity of 76.8, and 96.7\% of CUA-D traces end in the most common final goal observed for humans on the same task. CUA-SR shows weaker alignment: 76.7\% of its traces end in the most common human final goal, and it averages 1.17 goals per trace compared with about 1.5 for humans.

The L4 labels also reveal a difference that final task goals alone do not capture. Both agents use fewer distinct goals across their traces than humans (45 for CUA-D vs.\ 86 for SUs; 35 for CUA-SR vs.\ 88 for BLVUs). In particular, agents often omit auxiliary goals that appear in human traces, such as seeking help or closing extra windows and tabs. Thus, agents often match humans on the primary goal of the task, while differing in the additional goals humans pursue along the way.\

\begin{figure*}
    \centering
    \includegraphics[width=6in]{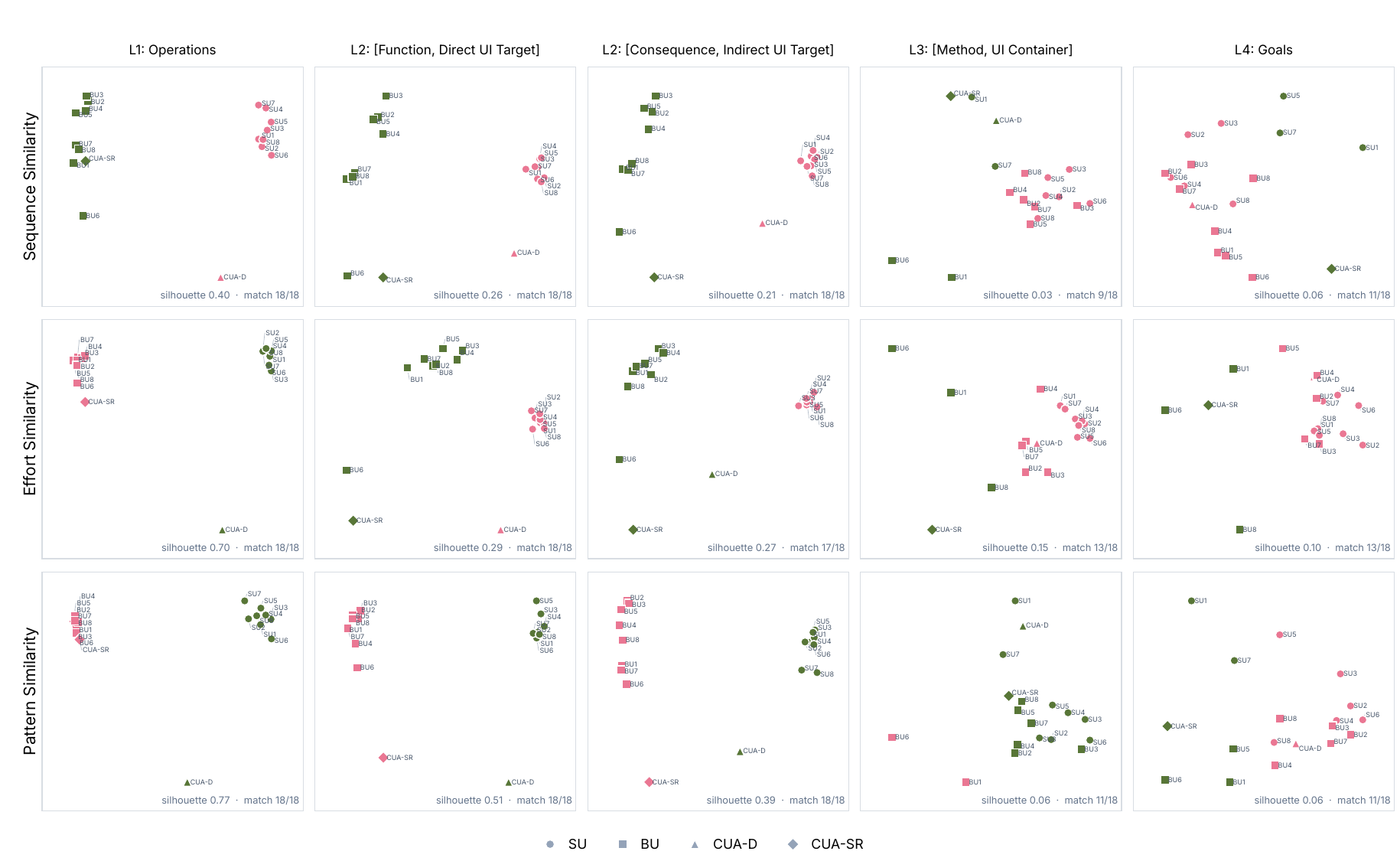}
    \caption{PCA projections of SU, BU, CUA-D, and CUA-SR traces across Sequence, Effort, and Pattern Similarity and \system{} abstraction levels. SU and BU are strongly separated at L1 and L2 but increasingly converge at L3 and L4, suggesting that group differences are greater in low-level execution than in higher-level methods and goals. Each panel reports the $k$-means ($k=2$) silhouette score and group-match count.}
    \Description{A 3-by-5 grid of PCA scatterplots compares sighted users (SU), blind users (BU), CUA-Default, and CUA-SR across three similarity metrics (rows: Sequence, Effort, Pattern) and five abstraction levels (columns: L1 Operations; two L2 representations; L3 Methods and UI Containers; L4 Goals). At L1, SU and BU form clearly separated clusters in all three rows, with the strongest separation for Effort and Pattern Similarity. Separation decreases across the two L2 columns and becomes weak at L3 and L4, where SU and BU points increasingly overlap. CUA-Default generally appears nearer the sighted-user cluster at lower levels, while CUA-SR generally appears nearer the blind-user cluster, though both agent conditions are often offset from the human points. Each panel also reports a silhouette score and the number of 18 traces whose cluster assignment matches the corresponding SU/BU group; these values are highest at L1 and generally decline toward L3 and L4.}
    \label{fig:pca}
\end{figure*}

\section{Discussion}

Our results show that the level at which interaction traces are compared matters: users and agents can differ substantially in their low-level actions while converging on similar methods and goals, and similar workflows can still differ in the effort required to carry them out. We discuss how \system{}'s linked abstraction levels can support applications that build on these distinctions, how the representation may generalize beyond computer use, and limitations of the current approach.

\subsection{Potential Applications of \system{}}

\noindent \textbf{\system{} can surface more efficient strategies from interaction traces.}
Our results suggest that \system{} could support systems that help users discover easier ways to complete computer tasks. Prior work has recommended commands and keyboard shortcuts to improve efficiency~\cite{cockburn2014supporting,matejka2009communitycommands,li2011design,malacria2013skillometers,wang2018leveraging}. \system{} offers a way to find such alternatives directly from interaction traces by comparing users who complete the same L3 Method but use different lower-level actions.

For example, three screen reader users all performed \texttt{convert} $\langle$file$\rangle$ in $\langle$spreadsheet$\rangle$, but took very different paths. One user pressed \texttt{F12} to open Save As directly and finished in 42 actions. Another opened the File menu with \texttt{Alt+F} and then pressed \texttt{a}, taking 70 actions. A third user moved through the menu item by item and took 100 actions, including 43 \texttt{navigate}$\rightarrow$\texttt{focus} actions. Because all three users were carrying out the same Method, \system{} makes it possible to identify the shorter interaction path and the specific shortcut that enabled it. We see a similar case for \texttt{export} $\langle$webpage$\rangle$ in $\langle$browser$\rangle$: one user opened the print dialog directly with \texttt{Ctrl+P}, while another reached it by navigating through browser controls.

This points to a broader use of semantic interaction traces: instead of recommending a shortcut in isolation, a system could first find users performing the same Method, identify which executions require fewer actions, and surface the strategy that produced that difference. Such recommendations would still need to account for context, since a trace alone cannot establish whether a shortcut is available, accessible, or appropriate in another user's current interface state. Future systems could combine \system{} with interface-state information and user preferences to decide when an observed strategy is actually useful to recommend.\\

\noindent \textbf{\system{} can flag potential usability issues from interaction traces.}
Our results also suggest that \system{} could help identify parts of an interface where users repeatedly encounter difficulty. Prior work has used interaction time, repeated actions, path deviations, and corrective actions as signals of usability problems~\cite{hong2001webquilt,shah2008event,akers2009undo,filosa2025travis}. \system{} adds a way to connect these signals to the part of the task where the difficulty occurs. By comparing the time and number of actions required for the same L3 Method across users, it can flag unusually costly Methods and then use the lower levels to show which interactions contributed to that cost.

\begin{figure}[t]
    \centering
    \includegraphics[width=3in]{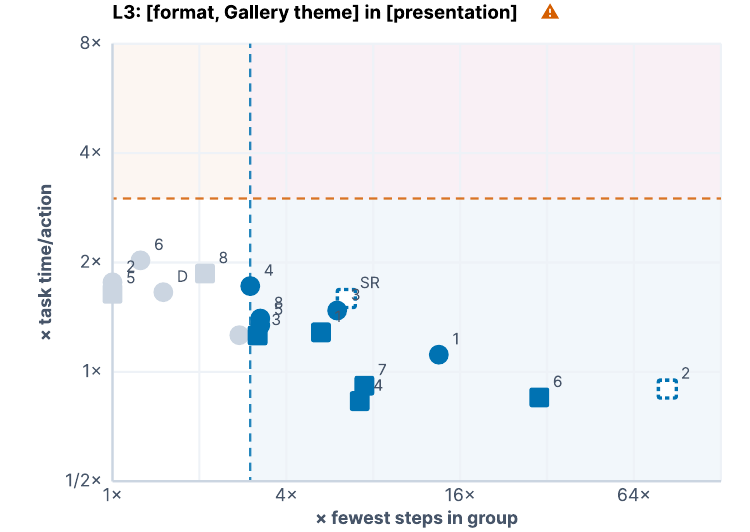}
    \caption{Difficulty in changing a PowerPoint theme. Each point represents one user's \texttt{format} Method. The horizontal axis shows actions relative to the most efficient successful user in the same group; the vertical axis shows time per action relative to the user's usual pace. Points beyond the thresholds are \emph{long-winded} or \emph{slow}. Most difficulty comes from requiring many actions rather than performing them slowly.}
    \Description{Scatterplot of L3 \texttt{[format, Gallery theme] in [presentation]} executions. The x-axis shows action count relative to the fewest successful actions within the same group (log scale, from 1× to 64×), and the y-axis shows time per action relative to each user’s task-level baseline (log scale, from 0.5× to 8×). Dashed thresholds mark unusually high action count and unusually slow execution. Most points fall below the time threshold but many lie to the right of the action-count threshold, indicating that applying the Gallery theme is more often difficult because it requires many extra actions than because individual actions are unusually slow.}
    \label{fig:usability-issue-fig}
\end{figure}

For example, in an A11y-CUA task that asked users to change a presentation to the \texttt{Gallery} theme, the \texttt{format} $\langle$Gallery theme$\rangle$ Method required unusually many actions for seven of nine blind users and five of nine sighted users (Fig.~\ref{fig:usability-issue-fig}). The interface presents themes in a horizontally scrolling gallery without a direct way to select \texttt{Gallery} by name. At L2, the flagged traces contain many repeated \texttt{navigate}$\rightarrow$\texttt{focus} interactions as users move through theme options to reach the desired theme. \system{} therefore identifies not only that this Method requires unusually high effort, but also that much of that effort comes from repeatedly navigating through the gallery.

More broadly, this could support usability analysis at scale. Rather than inspecting every trace manually, designers could use \system{} to find Methods that repeatedly require high effort across users, inspect the lower-level interactions associated with those Methods, and prioritize them for further study or redesign. Future systems could also combine these signals with information about the interface itself to distinguish difficulty caused by the interface from differences in user strategy, and to suggest possible changes such as adding direct selection, shortcuts, or more accessible navigation. These flags should be treated as candidates rather than automatic diagnoses: unusual time or action counts can indicate a usability problem, but still require inspection or follow-up testing to understand why it occurred. \\

\noindent\textbf{\system{} may help reduce identifiable information when sharing interaction traces.}
Interaction traces are useful for studying computer use, but they can also reveal who produced them. Prior work shows that keyboard, mouse, and GUI behavior can identify individual users and distinguish humans from agents~\cite{bailey2014user,wang2026fp,lugoloobi2026known}. Our results suggest that the level at which a trace is represented may affect how much of this identifying behavior remains.

We test this by predicting which of 18 trace sources produced a trace using the features available at each \system{} level, including label frequencies, action and time allocation, and similarity scores. Under leave-one-task-out evaluation, top-1 classification accuracy is 25.4\% at L1 and rises to 43.3\% at L2, showing that adding semantic information can initially make users \emph{more} distinguishable. At L3, however, accuracy drops to 13.0\%, closer to the 5.6\% expected by random chance. This drop appears across trace sources: classification falls from 83.3\% to 20.0\% for agents, 32.5\% to 8.3\% for SUs, and 44.2\% to 15.8\% for BLVUs.

This suggests a possible trade-off when sharing interaction data. L2 retains differences in how users navigate and interact with interfaces, which can make individual behavior easier to recognize. L3 instead keeps the Methods users perform while removing more of the lower-level detail that distinguishes how each person performs them. Method-level traces may therefore support some forms of behavioral analysis while exposing less identifying information than lower-level traces. However, lower classification accuracy does not by itself provide a privacy guarantee. Future work should test this with stronger identification attacks, larger populations, and combinations with other potentially identifying information before using semantic abstraction as a privacy mechanism.

\subsection{Generalizing Semantic Trace Abstraction Beyond Computer Use}

\system{} is designed for desktop GUI interaction, and its current vocabulary reflects that scope. For example, labels such as \texttt{keypress}, \texttt{navigate}, and \texttt{open} describe actions and outcomes that occur when people use desktop applications. Extending \system{} to a different interaction setting would therefore require new labels. What may generalize is not the vocabulary itself, but the idea of linking multiple levels that describe \emph{how an action is performed}, \emph{what it accomplishes}, \emph{which method it contributes to}, and \emph{which larger goal that method serves}.

This distinction becomes clearer for interfaces outside desktop computer use. In virtual reality, L1 might include controller movements, gaze, or gestures; L2 could describe effects such as selecting, grabbing, or moving a virtual object; L3 could represent methods such as placing or assembling objects; and L4 could capture the larger task goal. In robotics demonstrations, L1 could contain low-level robot motions, L2 their effects on objects, L3 manipulation methods such as grasping or placing, and L4 the task being completed~\cite{welschehold2019combined}. These domains would need different labels from \system{}, but could retain the same links between execution, outcome, method, and goal.

Programming and CAD are closer to our current setting because they are often performed through desktop interfaces, but their domain-specific behavior is not fully represented by our current vocabulary. A programming extension, for example, may need to distinguish methods such as debugging, refactoring, and testing rather than treating them only as generic editing or navigation. CAD traces may similarly need methods that capture domain-specific design operations rather than only the GUI actions used to invoke them~\cite{gopsill2016computer}. Extending \system{} would therefore mean keeping the abstraction structure while replacing or expanding the vocabulary to capture what is meaningful in that domain.

Our results suggest why this structure may be useful beyond desktop interaction. We repeatedly find that behaviors that look very different at a low level can accomplish the same Method or Goal, while important differences in effort remain visible in the lower levels. The same question arises in other domains: two people, or a person and a robot, may accomplish the same goal using very different actions. Comparing only the low-level actions can make equivalent strategies look unrelated, while comparing only the final goal hides how differently it was achieved. A multi-level representation preserves both. Future work could test whether the same design choices that were useful in \system{}---separating actions from their consequences, linking lower-level behavior to Methods, and retaining effort at each level---also support meaningful comparisons in VR, robotics, and other interactive domains.

\subsection{Limitations}

\system{} relies on information that is available across the interaction logs we study: keyboard and mouse actions and the available interface context around those actions~\cite{mohanbabu2026a11y,xie2026agentsynth,yang2026agentnet}. It does not require signals that are only available in some datasets, such as screen reader speech, screenshots aligned to every action, or accessibility-tree state. This allows the same pipeline to be applied across different logging setups, but it also limits what \system{} can infer. For example, a keypress may move screen reader focus even when that change is not recorded in the trace. Additional signals could help resolve such cases, and future work could test how much screenshots, accessibility-tree information, or screen reader output improve annotation accuracy.

L3 introduces another source of ambiguity because \system{} must decide both \emph{where one Method ends and another begins} and \emph{which Method those actions represent}. For example, when a user opens the Save As dialog and then changes a file's format, the action that opens Save As could be grouped with the preceding \texttt{open} Method or with the following \texttt{convert} Method. Moving this boundary changes which actions are grouped together and can therefore change the assigned Method. Specific Entities and L4 Goals are also open vocabulary: for example, two traces may describe the same goal using slightly different wording that must be mapped to the same concept. We canonicalize these labels, but some interactions can still reasonably be divided or described in more than one way. Our manual agreement results therefore show how consistently \system{} follows our annotation scheme, while recognizing that another reasonable annotation may sometimes differ.

The L3 Method and UI Container vocabularies were developed from A11y-CUA and then evaluated on A11y-CUA, AgentSynth, and AgentNet. They cover nearly all interactions in these datasets, but those datasets represent only a subset of computer use. Other workflows may introduce behavior that our current labels do not distinguish well. For example, creative work may involve repeated experimentation without a clear sequence of Methods, recovery may alternate between diagnosing and fixing a problem, and users may interleave several goals rather than completing one before starting another. Applying \system{} to these settings may therefore require adding or refining labels while keeping the same linked levels.

Finally, our similarity metrics capture three aspects of workflow behavior: the order of semantic labels, recurring patterns, and the effort spent at each level. They do not capture every meaningful difference between workflows. Prior work also compares loops, skipped steps, alternative paths~\cite{schreiber2025graph}, and how far a workflow progresses before failure~\cite{chung2026did}. Future work could extend \system{} to represent errors and recovery, transitions between applications, task success, and consistency across repeated attempts.

\section{Conclusion}
We presented \system{}, a hierarchical taxonomy for representing interaction traces across four levels of abstraction from input-level operations to high-level goals. \system{} uses an automated annotation pipeline to assign \system{} labels. The pipeline achieves strong agreement with human annotations and we demonstrate this on three datasets that record computer use interaction traces of humans and agents. \system{} also provides three similarity metrics (sequence, pattern, and effort) that can be computed at any level or combined across abstraction levels. We compared how sighted users, blind users, and computer use agents complete tasks using the A11y-CUA dataset. We show that \system{} replicates prior findings at lower abstraction levels while surfacing new findings at intermediate and higher abstraction levels.
We demonstrated \system{}'s use cases in improving novice users' expertise in computer use tasks by deriving efficient strategies from expert users, identifying potential usability issues in GUIs, and identifying users' behavioral fingerprints in computer use.
Together, \system{}'s hierarchical taxonomy and similarity metrics provide a unified framework for researchers analyzing and comparing interaction traces across multiple levels.


\bibliographystyle{ACM-Reference-Format}
\bibliography{references}

\appendix

\section{\system{} Prompts}
\label{sec:prompts}

\subsection{L2 Functions \& Consequences}
\label{sec:functions-prompt}

\begin{tcolorbox}[
  breakable,
  colback=white,
  colframe=black!35,
  arc=5pt,
  boxrule=0.4pt,
  left=8pt,
  right=8pt,
  top=7pt,
  bottom=7pt
]
Infer L2 labels for each action in the UI interaction trace. L2 structure: {Function, Direct UI Target} $\rightarrow$ {Consequence, Indirect UI Target}. \\

\noindent Function shows what the operation did (already determined for each action in the input). Direct UI Target is the element on which the action occurs. Choose the most specific directly interacted-with element supported by the trace. For example, use cell rather than table and list item rather than list for Direct UI Target. Consequence is the direct result of the action. Indirect UI Target is the affected UI object on which change is observed because of the consequence of the action. \\

\noindent Examples: a "select" Function on a "list item" (Direct UI Target) representing a file, where the action opens that file, gives the Consequence "open" on Indirect UI Target "file". A "drag" Function on "list item" (Direct UI Target) representing a file, dropped into a folder, gives the Consequence "move" and Indirect UI Target "folder". \\

\noindent Instructions for determining the L2 output for each operation: 
\begin{itemize}
    \item The direct and indirect target may be the same or different. For direct target, choose the specific UI element on which action occurs and not its container (indirect UI target).
    \item For point actions, infer the direct and indirect UI target from the destination, using the following action only when needed. For keyboard actions, use the focused element as the Direct UI Target.
    \item Use the given Function, lower-level labels, and raw action details (event, keys, application, window\_title, target, focused\_element, and nearby actions) to infer the remaining labels.
    \item For each action, output only index, direct\_ui\_target, consequence, indirect\_ui\_target, rationale, confidence (0–1), and a short plain-language description.
    \item Output a JSON array with exactly \texttt{actionCount} entries, one per action, sorted by index. Do not merge or omit actions.
\end{itemize}
\end{tcolorbox}

\subsection{L3 Methods}
\label{sec:methods-prompts}

\begin{tcolorbox}[
  breakable,
  colback=white,
  colframe=black!35,
  arc=5pt,
  boxrule=0.4pt,
  left=8pt,
  right=8pt,
  top=7pt,
  bottom=7pt
]
Infer a canonical segmentation and L3 semantic label for every action in the UI interaction trace. L3 structure: \texttt{[Method, Specific Entity] in \{UI Container\}}. \\

\noindent Your first task is segmentation. Divide the complete trace into contiguous task-level procedures, then assign exactly one L3 label to every procedure. All actions within a procedure must receive the exact same \texttt{Method}, \texttt{Specific Entity}, and \texttt{UI Container}. Start a new segment only when the task-level objective or completed outcome changes. Also segment when the active application changes on desktop or when the webpage changes in a browser. Do not create boundaries for changes in event type, menus or dialogs, navigation, verification, corrections, retries, pauses, or interaction technique. Setup, recovery, and confirmation actions inherit the procedure they support. \\

\noindent Reference example: use the following manually annotated trace as a reference for the intended L3 segmentation granularity. \\

\noindent \textbf{Task:} ``Add the video \texttt{cat.mp4} to \texttt{family.pptx}, both in \texttt{Document > My-Home} folder.'' \\

\begin{itemize}
    \item {Segment 1 - Navigate to the task folder.} Open File Explorer and navigate through folders to \texttt{My-Home}. \\
    L3: \texttt{[open, My-Home folder] in \{file explorer\}}

    \item {Segment 2 - Copy the video.} Select \texttt{cat.mp4} and copy it to the clipboard. \\
    L3: \texttt{[export, cat.mp4] in \{file explorer\}}

    \item Segment 3 - Open the presentation. Select \texttt{family.pptx} and open it, the presentation interface becomes active. \\
    L3: \texttt{[open, family.pptx] in \{presentation\}}

    \item Segment 4 - Add the video to the presentation. Focus the presentation canvas and paste the copied video. \\
    L3: \texttt{[add, cat.mp4] in \{presentation\}}
\end{itemize}

\noindent \texttt{Method} represents the subgoal that an action contributes to, not what the individual action immediately does. For example, opening a font menu, choosing a size, and confirming it are all part of \texttt{format} text in a text document, rather than separate \texttt{open}, \texttt{select}, and \texttt{confirm} Methods. \texttt{Specific Entity} is the shortest grounded name of the task-relevant item on which the Method is performed. Use the visible or referenced name when available (\textit{e.g.}, \texttt{My-Home}); otherwise, use a concise descriptor. \texttt{Specific Entity} is open vocabulary. \texttt{UI Container} is the enclosing document or application type in which the procedure occurs. When multiple UI Container labels could apply, choose the container corresponding most directly to the content or workspace being interacted with (\textit{e.g.}, spreadsheet over website, text document over browser, file explorer over desktop). Choose the closest matching label from the allowed Method and UI Container lists, and use \texttt{unknown} when no listed label applies. \\

\noindent Instructions for determining the L3 output for each operation:
\begin{itemize}
    \item A reader should be able to read the three fields together as one coherent procedure: \texttt{[Method, Specific Entity] in \{UI Container\}} (\textit{e.g.}, ``edit, Introduction, in text document'').
    \item Segment the complete trace first, then assign one L3 label to each segment. Every action within the same segment must receive identical \texttt{method}, \texttt{specific\_entity}, and \texttt{ui\_container} values.
    \item Use lower-level labels and raw action details (\texttt{event}, \texttt{keys}, \texttt{application}, \texttt{window\_title}, \texttt{target}, \texttt{focused\_element}, and nearby actions) to infer segmentation and labels.
    \item The L3 label should fit naturally with the preceding and following segments in the overall interaction flow.
    \item For each action, output only \texttt{index}, \texttt{method}, \texttt{specific\_entity}, \texttt{ui\_container}, \texttt{rationale}, \texttt{confidence} (0--1), and a short plain-language \texttt{description}.
    \item Output a JSON array with exactly \texttt{actionCount} entries, one per action, sorted by index. Do not merge or omit actions.
\end{itemize}
\end{tcolorbox}

\subsection{L4 Goals}
\label{sec:goals-prompts}

\begin{tcolorbox}[
  breakable,
  colback=white,
  colframe=black!35,
  arc=5pt,
  boxrule=0.4pt,
  left=8pt,
  right=8pt,
  top=7pt,
  bottom=7pt
]
Infer L4 semantic labels for actions in the UI interaction trace.\\
\noindent Goal is a short, task-level phrase that names the user’s objective achieved by a group of L3 methods, describing the outcome of the high-level goals. A goal must normally contain multiple L3 steps rather than restating a single L3 Method. \\
\noindent Instructions for determining the L4 output for each operation: 
\begin{itemize}
    \item Use the lower-level labels and lower-level action details (event, keys, application, window\_title, target, focused\_element, and nearby actions) for grounding.
    \item Group contiguous L3 steps that jointly serve the same objective.
    \item Every L3 step must belong to exactly one Goal; groups cannot overlap or skip intervening steps.
    \item For each output goal, include \texttt{l3StepIndices} (all covered L3 step indices, in order), rationale, and confidence (0–1).
\end{itemize}
\end{tcolorbox}

\newpage
\onecolumn
\section{\system{} Taxonomy Vocabulary}
\label{sec:vocabulary}

\subsection{Consequences (L2)}
\label{sec:consequences}

\begin{table}[H]
\centering
\setlength{\tabcolsep}{6pt}
\begin{tabular}{@{}l p{0.88\textwidth}@{}}
\toprule
\textbf{Label} & \textbf{Definition} \\
\midrule
\texttt{cancel} & Stop or abandon a pending operation without applying or committing its intended changes. \\
\texttt{deselect} & Remove an existing selection. \\
\texttt{toggle} & Switch a control between two states, such as on/off. \\
\texttt{zoom} & Change the magnification of the current view. \\
\texttt{cut} & Remove selected content and place it on the clipboard. \\
\texttt{copy} & Place selected content on the clipboard without removing it. \\
\texttt{paste} & Insert clipboard content at the current target. \\
\texttt{delete} & Remove content, an item, or an object. \\
\texttt{undo} & Reverse the most recent change. \\
\texttt{redo} & Reapply a previously undone change. \\
\texttt{close} & Dismiss an open tab, menu, dialog, view, or window from current view. \\
\texttt{maximize} & Expand a window to fill the available screen area. \\
\texttt{focus} & Keyboard/input focus moves to a target (e.g. clicking into a field, tabbing to a control) but no new UI surface appears. \\
\texttt{save} & Save the current document, file, or state. \\
\texttt{create} & Make a new item or UI object. \\
\texttt{select all} & Select all applicable content or items in the current scope. \\
\texttt{minimize} & Hide a window while keeping it open. \\
\texttt{open} & A new UI surface becomes visible that wasn't before, e.g. a menu drops down, a dialog appears, a panel expands. \\
\texttt{highlight} & Content is selected or emphasised across a range, usually by shift+arrow or dragging. \\
\texttt{move} & An object, window, or content is itself relocated spatially, by dragging. \\
\texttt{text input} & Add or edit textual content. \\
\texttt{autofill} & Automatically populate one or more input fields with stored, suggested information, rather than entering the values manually. \\
\texttt{hover} & Move or hold the pointer over an element without activating it, whether or not a visual change appears. \\
\texttt{pan} & The visible viewport shifts across a larger canvas or scrollable area, by scrolling. \\
\texttt{confirm} & Finalizes and closes a pending choice or flow, e.g. clicking OK/submit on a dialog. \\
\texttt{announce} & A bare Alt tapped on its own while a screen reader is running, where the screen reader intercepts the modifier press and re-speaks the current item, state, or status; ordinary navigation keys remain navigate or focus even when speech follows as a side effect. \\
\texttt{mute} & A bare Ctrl tapped on its own while a screen reader is running, where the screen reader intercepts the modifier press and silences the speech currently being spoken. \\
\texttt{none} & Action had no effect, e.g. a lone modifier key press. \\
\texttt{unknown} & None of the other consequence options clearly fit this action. \\
\bottomrule
\end{tabular}
\caption{L2 \textbf{Consequence} vocabulary. Consequence identifies the immediate result of an action on the interface or application state. The vocabulary contains 28 labels plus \texttt{unknown}; all 28 labels occur in A11y-CUA.}
\Description{The table lists the L2 Consequence labels used by \system{} to describe what immediately happens after an action, such as moving focus, entering text, opening an interface element, or changing application state. All 28 defined Consequence labels are observed in A11y-CUA; \texttt{unknown} is used when the trace does not provide enough information to determine the result.}
\label{tab:vocab-conseq}
\end{table}

\subsection{UI Targets (L2)}
\label{sec:ui-targets}
\begin{table}[H]
\centering
\setlength{\tabcolsep}{6pt}
\begin{tabular}{@{}l p{0.85\textwidth}@{}}
\toprule
\textbf{Label} & \textbf{Definition} \\
\midrule
\texttt{button} & A clickable UI control (e.g. toolbar button) that triggers an action when selected. \\
\texttt{radio button} & A single selectable option in a set of mutually exclusive choices. \\
\texttt{link} & A specific, identifiable clickable element (text, hyperlink, or image) that triggers navigation to a different location. \\
\texttt{search box} & An input field for entering a query or URL to search with a webpage or application. \\
\texttt{slider} & A draggable control for setting a numeric value within a range. \\
\texttt{date picker}$^{\dagger}$ & A calendar widget for selecting a specific date. \\
\texttt{color picker} & A widget for selecting a color from a palette, spectrum, or input value. \\
\texttt{checkbox} & A binary toggle that can be checked or unchecked independently of other options. \\
\texttt{menu item} & A single actionable option inside an open menu, dropdown, or context menu, e.g., File menu $\rightarrow$ Save As is the menu item. \\
\texttt{form} & A group of related fields and controls used to enter or submit information. \\
\texttt{table} & A 2-dimensional structure with data organized into a grid of rows and columns. \\
\texttt{cell} & A specific unit within a table, defined by a row and column position. \\
\texttt{column} & A vertical series of cells in a table. \\
\texttt{row} & A horizontal series of cells in a table. \\
\texttt{window} & A discrete container that displays an application's contents in a resizable frame. \\
\texttt{scrollbar} & A draggable control that moves the viewport through content that extends beyond the visible area. \\
\texttt{webpage} & A page's general content or background, e.g. clicking whitespace. \\
\texttt{text area} & An editable field or area where the user can enter or modify written content other than search box. \\
\texttt{image} & A static photo, illustration, or graphic viewed or placed as content, e.g. a thumbnail or inserted picture. \\
\texttt{video} & A dynamic visual media element that can be played, paused, or scrubbed. \\
\texttt{audio}$^{\dagger}$ & A media element containing only sound content. \\
\texttt{canvas} & An editable workspace where visual content is drawn or manipulated, e.g. a chart canvas or drawing tool's working area. \\
\texttt{desktop} & The main system workspace containing windows, icons, and other items. \\
\texttt{list} & A grouped collection of related items or options. \\
\texttt{list item} & A single entry within a list (e.g. file list in a folder, icons in taskbar). \\
\texttt{caret} & Active cursor position where text insertion happens. \\
\texttt{command} & A text or voice instruction issued to execute an action, e.g. via a command line. \\
\texttt{clipboard}$^{\dagger}$ & Temporary system storage used by cut, copy, and paste operations. \\
\texttt{autofill} & Automatically populate one or more input fields with suggested information by moving the caret, rather than entering the values manually. \\
\texttt{unknown} & None of the other direct UI target options clearly fit this action. \\
\midrule

\texttt{file} & A UI item containing data, content, or executable information. \\
\texttt{context menu} & A menu opened for a specific item or location, usually by right-clicking, containing actions for that context; unlike a regular menu, it is tied to the item or location where it was opened. \\
\texttt{dialog box} & A temporary modal that appears over the current interface to request input, confirm an action, or display a message before the user can continue. \\
\texttt{item card} & A bounded card element displaying details for a single item in a collection, feed, or grid layout. \\
\texttt{folder} & A container used to organize files and subfolders. \\
\texttt{application} & A desktop application excluding the browser. \\
\texttt{none} & No UI object was affected, because the action produced no change in the interface; applies whenever the Consequence is none, mute, or announce. Distinct from unknown, which means an object was affected but could not be identified. \\

\bottomrule
\end{tabular}
\caption{L2 \textbf{Direct UI Target} vocabulary. Direct UI Target identifies the interface element on which an action is performed. \system{} assigns the most specific element supported by the trace (e.g., \texttt{cell} rather than \texttt{table}). The top section lists all Direct UI Target labels. The bottom section extends this vocabulary with additional labels that can be used only as Indirect UI Targets. The Direct UI Target vocabulary contains 29 labels plus \texttt{unknown}; 26 of the 29 labels occur in A11y-CUA. $^{\dagger}$ indicates labels not observed in A11y-CUA.}
\Description{The table shows the interface-element vocabulary used at L2. The top section contains all labels that can be assigned as Direct UI Targets, describing the element an action is performed on, such as a button, cell, text field, or menu item. \system{} uses the most specific target supported by the trace. The bottom section adds labels that are available only for Indirect UI Targets, which describe elements affected by an action even when the action is performed on something else. Labels marked with $^{\dagger}$ are defined in the vocabulary but were not observed in A11y-CUA.}
\label{tab:vocab-direct}
\end{table}

\subsection{Methods (L3)}
\label{sec:methods}
\begin{table}[H]
\centering
\setlength{\tabcolsep}{6pt}
\begin{tabular}{@{}l p{0.85\textwidth}@{}}
\toprule
\textbf{Label} & \textbf{Definition} \\
\midrule
\texttt{search} & Locate a specific item or information by entering or submitting a query through a search feature. \\
\texttt{export} & Save a separate output copy of existing content to a destination; convert applies instead when changing the file format is itself the objective. \\
\texttt{open} & Activate a specific content item, such as a file or document, so its contents become available for use; reaching a website through the browser is not open. \\
\texttt{verify} & Examine content of a UI item to check its information without modifying it, e.g. verifying product added to cart, changes to a document. \\
\texttt{edit} & Change only the textual contents of a UI item (e.g. changing text in a document, a filename, or form response) or change an existing numeric or slider-style setting of a UI item to a new value via an opened settings panel. \\
\texttt{format} & Change how existing textual and visual content is presented, without changing its substantive meaning, such as font, alignment, color, style, or layout. \\
\texttt{move} & An object, window, or content is itself relocated spatially, by dragging. \\
\texttt{dismiss} & Close or minimize a window, transient popup, context menu, or confirmation without acting on it further. \\
\texttt{convert} & Transform an existing file from one file format into another while preserving its underlying content. \\
\texttt{zoom}$^{\dagger}$ & Change magnification level of the content in current view. \\
\texttt{checkout}$^{\dagger}$ & Complete a purchase flow, e.g. confirming payment or submitting an order. \\
\texttt{add} & Add a brand-new object into a document via a ribbon or menu command or create a UI item via an ``add'' button. \\
\texttt{filter} & Apply a selection criterion on an already available data range or list to narrow it down. \\
\texttt{unknown} & None of the other method options clearly fit this action. \\
\bottomrule
\end{tabular}
\caption{L3 \textbf{Method} vocabulary. Method describes how a subgoal is carried out. We first open coded 53 Method terms during an annotation pilot, then consolidated them into 13 labels that were used consistently across independent annotation runs. The vocabulary contains 13 labels plus \texttt{unknown}; 11 of the 13 labels occur in A11y-CUA. $^{\dagger}$ indicates labels not observed in A11y-CUA.}
\Description{The table lists the L3 Method labels used by \system{} to describe how a user or agent carries out a subgoal, such as \texttt{open}, \texttt{edit}, \texttt{search}, or \texttt{verify}. The vocabulary was developed from 53 Method terms identified during an annotation pilot and consolidated into 13 labels that could be applied consistently across traces. Eleven labels appear in A11y-CUA; labels marked with $^{\dagger}$ are part of the vocabulary but were not observed in that dataset. \texttt{unknown} is used when the trace does not provide enough information to assign a Method.}
\label{tab:vocab-method}
\end{table}

\subsection{UI Containers (L3)}
\label{sec:ui-containers}
\begin{table}[H]
\centering
\setlength{\tabcolsep}{6pt}
\begin{tabular}{@{}l p{0.85\textwidth}@{}}
\toprule
\textbf{Label} & \textbf{Definition} \\
\midrule
\texttt{browser} & The browser's own interface, such as tabs, address bar, bookmarks, navigation controls, downloads, or browser settings; excludes the content of loaded webpages. \\
\texttt{website} & The content and interface of a loaded website or web app. \\
\texttt{image editor} & An application used to view, create, or edit images or graphics. \\
\texttt{media player} & An application used to play or control audio or video media. \\
\texttt{code editor} & An application for writing, editing, and running source code. \\
\texttt{canvas}$^{\dagger}$ & An interactive area where visual content is drawn or manipulated. \\
\texttt{spreadsheet} & A worksheet containing data organized into rows and columns. \\
\texttt{text document} & An application to view or edit written text in a document. \\
\texttt{email client}$^{\dagger}$ & An interface for reading, composing, organizing, or managing email. \\
\texttt{presentation} & An application used to view, create, edit, or present slides in a slide-deck. \\
\texttt{file explorer} & A system interface for browsing and managing files and folders. \\
\texttt{desktop} & The operating system workspace containing windows, icons, and system controls. \\
\texttt{clock} & An application for viewing time and managing alarms, timers, stopwatches, or world clocks. \\
\texttt{settings} & A system-level interface for changing preferences, permissions, accessibility options, or other configuration. \\
\texttt{application} & A native desktop application whose interface doesn't match any other listed UI container. \\
\texttt{unknown} & None of the other UI container options fit this action. \\
\bottomrule
\end{tabular}
\caption{L3 \textbf{UI Container} vocabulary. UI Container identifies the application area in which a subgoal is carried out, such as a dialog, menu, toolbar, or document. The vocabulary contains 15 labels plus \texttt{unknown}; 13 of the 15 labels occur in A11y-CUA. $^{\dagger}$ indicates labels not observed in A11y-CUA.}
\Description{The table lists the L3 UI Container labels used by \system{} to describe where a Method is carried out within an application. These labels capture the broader interface area containing the interaction, such as a dialog, menu, toolbar, or document. Thirteen of the 15 defined labels appear in A11y-CUA; labels marked with $^{\dagger}$ are part of the vocabulary but were not observed in that dataset. \texttt{unknown} is used when the trace does not provide enough information to determine the container.}
\label{tab:vocab-container}
\end{table}

\section{Within- and Across Group Similarities Across \system{} Abstraction Levels}

\begin{table}[H]
\centering
\small
\setlength{\tabcolsep}{6pt}
\renewcommand{\arraystretch}{1.1}

\begin{tabular}{@{}lccc@{}}
\toprule
\textbf{Level} & \textbf{Sequence} & \textbf{Effort} & \textbf{Pattern} \\
\midrule
L1  & 0.40 & 0.70 & 0.77 \\
L2a & 0.26 & 0.29 & 0.51 \\
L2b & 0.21 & 0.27 & 0.39 \\
L3  & 0.03 & 0.15 & 0.06 \\
L4  & 0.06 & 0.10 & 0.06 \\
\bottomrule
\end{tabular}

\caption{
Silhouette scores for $k$-means clustering ($k=2$) across \system{} abstraction levels and similarity metrics. Higher scores indicate stronger separation between the two clusters; separation is strongest at L1 and decreases at higher abstraction levels.
}
\Description{Table reporting silhouette scores for two-cluster \(k\)-means across MAP abstraction levels and the three similarity metrics. Separation is strongest at L1, especially for Pattern (0.77) and Effort (0.70), declines across the two L2 representations, and becomes weak at L3 and L4, where all scores are at or below 0.15. This indicates that group structure is most pronounced in low-level interaction and largely disappears at higher semantic levels.}
\label{tab:silhouette}
\end{table}

\begin{table}[H]
\centering
\small
\setlength{\tabcolsep}{2pt}
\renewcommand{\arraystretch}{1.12}

\begin{tabular}{@{}llcccccccc@{}}
\toprule
\textbf{Metric} &
\textbf{Level} &
\textbf{Within SU} &
\textbf{Within BU} &
\textbf{SU--CUA-D} &
\textbf{SU--BU} &
\textbf{SU--CUA-SR} &
\textbf{CUA-D--BU} &
\textbf{CUA-D--CUA-SR} &
\textbf{BU--CUA-SR} \\
\midrule

\multirow{5}{*}{Sequence}
& L1
& 52.90 {\scriptsize (3.31)}
& 46.60 {\scriptsize (5.32)}
& 34.73 {\scriptsize (3.23)}
& 12.79 {\scriptsize (2.98)}
& 11.70 {\scriptsize (1.96)}
& 13.77 {\scriptsize (2.51)}
& 14.73
& 42.93 {\scriptsize (2.19)} \\

& L2a
& 49.21 {\scriptsize (2.15)}
& 36.88 {\scriptsize (4.79)}
& 38.82 {\scriptsize (2.20)}
& 21.17 {\scriptsize (4.07)}
& 18.98 {\scriptsize (1.22)}
& 16.37 {\scriptsize (3.07)}
& 16.92
& 27.21 {\scriptsize (1.80)} \\

& L2b
& 41.62 {\scriptsize (2.04)}
& 30.43 {\scriptsize (3.80)}
& 30.11 {\scriptsize (2.45)}
& 16.39 {\scriptsize (3.20)}
& 14.53 {\scriptsize (1.14)}
& 15.88 {\scriptsize (3.19)}
& 19.05
& 22.58 {\scriptsize (1.25)} \\

& L3
& 71.95 {\scriptsize (6.12)}
& 67.48 {\scriptsize (5.67)}
& 68.91 {\scriptsize (5.15)}
& 68.77 {\scriptsize (6.76)}
& 62.85 {\scriptsize (3.70)}
& 64.38 {\scriptsize (4.60)}
& 66.22
& 61.42 {\scriptsize (4.39)} \\

& L4
& 72.21 {\scriptsize (6.71)}
& 73.76 {\scriptsize (5.44)}
& 74.44 {\scriptsize (6.58)}
& 70.66 {\scriptsize (6.46)}
& 62.29 {\scriptsize (4.66)}
& 76.46 {\scriptsize (4.24)}
& 66.67
& 64.44 {\scriptsize (4.53)} \\

\midrule

\multirow{5}{*}{Effort}
& L1
& 72.76 {\scriptsize (5.12)}
& 86.50 {\scriptsize (4.15)}
& 25.57 {\scriptsize (4.67)}
& 10.71 {\scriptsize (4.03)}
& 3.15 {\scriptsize (1.90)}
& 5.91 {\scriptsize (1.03)}
& 0.92
& 73.25 {\scriptsize (4.02)} \\

& L2a
& 69.42 {\scriptsize (2.53)}
& 46.49 {\scriptsize (8.67)}
& 51.65 {\scriptsize (2.80)}
& 38.84 {\scriptsize (10.18)}
& 16.00 {\scriptsize (2.80)}
& 22.80 {\scriptsize (6.92)}
& 10.02
& 27.87 {\scriptsize (3.77)} \\

& L2b
& 55.23 {\scriptsize (3.20)}
& 29.17 {\scriptsize (7.10)}
& 24.41 {\scriptsize (2.49)}
& 14.61 {\scriptsize (5.35)}
& 5.08 {\scriptsize (0.93)}
& 11.47 {\scriptsize (3.59)}
& 6.75
& 11.58 {\scriptsize (2.90)} \\

& L3
& 78.12 {\scriptsize (4.02)}
& 53.57 {\scriptsize (9.33)}
& 68.50 {\scriptsize (2.13)}
& 59.91 {\scriptsize (10.93)}
& 43.96 {\scriptsize (3.87)}
& 54.31 {\scriptsize (8.25)}
& 51.44
& 45.54 {\scriptsize (6.56)} \\

& L4
& 64.80 {\scriptsize (6.58)}
& 54.42 {\scriptsize (9.80)}
& 62.23 {\scriptsize (6.27)}
& 54.33 {\scriptsize (10.07)}
& 42.91 {\scriptsize (4.76)}
& 59.52 {\scriptsize (8.21)}
& 50.54
& 47.85 {\scriptsize (6.59)} \\

\midrule

\multirow{5}{*}{Pattern}
& L1
& 86.20 {\scriptsize (3.32)}
& 93.82 {\scriptsize (2.45)}
& 42.83 {\scriptsize (4.51)}
& 33.15 {\scriptsize (4.97)}
& 30.93 {\scriptsize (4.59)}
& 38.83 {\scriptsize (1.26)}
& 39.44
& 88.69 {\scriptsize (1.57)} \\

& L2a
& 72.07 {\scriptsize (3.43)}
& 76.85 {\scriptsize (4.23)}
& 51.12 {\scriptsize (4.22)}
& 41.11 {\scriptsize (3.08)}
& 33.91 {\scriptsize (2.98)}
& 34.39 {\scriptsize (2.30)}
& 33.04
& 57.84 {\scriptsize (2.81)} \\

& L2b
& 61.09 {\scriptsize (3.30)}
& 58.82 {\scriptsize (6.14)}
& 41.50 {\scriptsize (4.68)}
& 27.65 {\scriptsize (3.66)}
& 23.99 {\scriptsize (3.42)}
& 32.38 {\scriptsize (3.88)}
& 36.97
& 44.16 {\scriptsize (1.43)} \\

& L3
& 76.76 {\scriptsize (5.42)}
& 71.41 {\scriptsize (5.35)}
& 73.54 {\scriptsize (5.00)}
& 72.51 {\scriptsize (6.26)}
& 72.16 {\scriptsize (3.65)}
& 67.89 {\scriptsize (5.15)}
& 71.85
& 70.68 {\scriptsize (3.76)} \\

& L4
& 77.46 {\scriptsize (6.09)}
& 76.39 {\scriptsize (5.19)}
& 77.47 {\scriptsize (4.69)}
& 74.74 {\scriptsize (6.50)}
& 68.30 {\scriptsize (4.03)}
& 78.79 {\scriptsize (5.07)}
& 72.05
& 70.55 {\scriptsize (4.15)} \\

\bottomrule
\end{tabular}

\caption{
Within- and between-group Sequence, Effort, and Pattern Similarity across \system{} abstraction levels. Values report mean similarity with standard deviation in parentheses. CUA-D--CUA-SR contains a single pair and therefore has no standard deviation.
}
\Description{Table reporting within- and between-group Sequence, Effort, and Pattern Similarity across MAP abstraction levels for sighted users (SU), blind users (BU), CUA-D, and CUA-SR. At L1 and L2, within-group similarities are generally much higher than SU–BU similarities, especially for Effort and Pattern, indicating strong group differences in low-level execution. At L3 and L4, Sequence and Pattern similarities become much more similar across within- and between-group comparisons, showing convergence in higher-level Methods and Goals. Effort remains more variable, particularly for BU and CUA-SR comparisons. CUA-D is generally closer to SU and CUA-SR to BU at lower levels.}
\label{tab:similarity}
\end{table}

\end{document}